\documentclass[journal,twoside]{IEEEtran}

\usepackage{amsmath,amssymb,amsfonts}
\usepackage{mathtools}
\usepackage{bm}
\usepackage{graphicx}
\usepackage{cite}
\usepackage{algorithmic}
\usepackage{algorithm}
\usepackage{textcomp}
\usepackage{stfloats}
\usepackage{subcaption}

\graphicspath{{fig/}}

\newlength{\myfigwidth}
\newlength{\mysubfigwidth}

\newcommand{\ao}{\ensuremath{a^\circ}}

\newcommand{\bHh}{\ensuremath{\hat{\bm{H}}}}
\newcommand{\bW}{\ensuremath{\bm{W}}}

\newcommand{\bw}{\ensuremath{\bm{w}}}

\newcommand{\bv}{\ensuremath{\bm{v}}}
\newcommand{\bV}{\ensuremath{\bm{V}}}

\newcommand{\bhh}{\ensuremath{\hat{\bm{h}}}}
\newcommand{\bhb}{\ensuremath{\bar{\bm{h}}}}

\newcommand{\bh}{\ensuremath{\bm{h}}}

\newcommand{\brM}{\ensuremath{\bar{M}}}
\newcommand{\brB}{\ensuremath{\bar{B}}}
\newcommand{\brK}{\ensuremath{\bar{K}}}

\newcommand{\Btot}{\ensuremath{B_{\text{tot}}}}

\newcommand{\gth}{\ensuremath{\gamma_{\text{th}}}}

\newcommand{\hBtot}{\ensuremath{\hat{B}_\text{tot}}}

\newcommand{\mo}{\ensuremath{m^\circ}}

\newcommand{\Ptot}{\ensuremath{P_{\text{tot}}}}

\newtheorem{prop}{Proposition}

\newcounter{MYtempeqncnt}

\makeatletter
\def\ps@IEEEtitlepagestyle{%
  \def\@oddfoot{\mycopyrightnotice}%
  \def\@oddhead{\hbox{}\@IEEEheaderstyle\leftmark\hfil\thepage}\relax
  \def\@evenhead{\@IEEEheaderstyle\thepage\hfil\leftmark\hbox{}}\relax
  \def\@evenfoot{}%
}
\def\mycopyrightnotice{%
  \begin{minipage}{\textwidth}
    \centering \scriptsize \copyright~2023 IEEE. Personal
    use of this material is permitted.  Permission from IEEE must be
    obtained for all other uses, in any current or future media,
    including reprinting/republishing this material for advertising or
    promotional purposes, creating new collective works, for resale or
    redistribution to servers or lists, or reuse of any copyrighted
    component of this work in other works.
  \end{minipage}
}
\makeatother

\begin{document}

\title{Optimal Transmit Power and Channel-Information Bit Allocation
  With Zeroforcing Beamforming in MIMO-NOMA and MIMO-OMA Downlinks}

\author{Kritsada Mamat and Wiroonsak
  Santipach,~\IEEEmembership{Senior~Member,~IEEE}%
\thanks{This work was supported by the Program Management Unit for
    Human Resources and Institutional Development, Research and
    Innovation through the Office of National Higher Education Science
    Research and Innovation Policy Council (NXPO) under Grant
    B05F630035. {\it (Corresponding author: Wiroonsak Santipach.)}}
\thanks{K. Mamat is with the Department of Electronic Engineering
  Technology; College of Industrial Technology; King Mongkut's
  University of Technology North Bangkok, Bangkok, 10800, Thailand
  (email: kritsada.m@cit.kmutnb.ac.th).}
\thanks{W. Santipach is with the Department of Electrical Engineering;
  Faculty of Engineering; Kasetsart University, Bangkok, 10900,
  Thailand (email: wiroonsak.s@ku.ac.th).}}

\markboth{Accepted by IEEE Transactions on Communications, 2023}%
         {Mamat and Santipach: Optimal Tx Power and
           Channel-Info Bit Alloc. With ZF
           BF in MIMO-NOMA and MIMO-OMA Downlinks}

\maketitle

\begin{abstract}
In downlink, a base station (BS) with multiple transmit antennas
applies zeroforcing beamforming to transmit to single-antenna mobile
users in a cell.  We propose the schemes that optimize transmit power
and the number of bits for channel direction information (CDI) for all
users to achieve the max-min signal-to-interference plus noise ratio
(SINR) fairness. The optimal allocation can be obtained by a geometric
program for both non-orthogonal multiple access (NOMA) and orthogonal
multiple access (OMA).  For NOMA, 2 users with highly correlated
channels are paired and share the same transmit beamforming.  In some
small total-CDI rate regimes, we show that NOMA can outperform OMA by
as much as 3 dB.  The performance gain over OMA increases when the
correlation-coefficient threshold for user pairing is set higher.  To
reduce computational complexity, we propose to allocate transmit power
and CDI rate to groups of multiple users instead of individual users.
The user grouping scheme is based on K-means over the user SINR.  We
also propose a progressive filling scheme that performs close to the
optimum, but can reduce the computation time by almost 3 orders of
magnitude in some numerical examples.
\end{abstract}


\section{Introduction}

Combining multiple-input multiple-output (MIMO) with non-orthogonal
multiple access (NOMA) can support more users and achieve higher
spectral efficiency in wireless networks than that with orthogonal
multiple access (OMA)~\cite[and references therein]{dai2018, mzeng17,
  zhang17,liu16}. Users in power-domain NOMA may share the same
time/frequency/spatial resource, but are distinguished by proper
transmit-power allocation and successive interference cancellation
(SIC)~\cite{dai2018}. However, the gain from NOMA is traded with
additional complexity from superposition coding, power allocation,
user clustering, and SIC~\cite{nain17,im19,bisen21}.

In downlink, a base station (BS) with multiple transmit antennas
employs zeroforcing beamforming to transmit message signals to all
users in a cell.  Zeroforcing beamforming from the BS nulls out all
multi-user interference when the signal reaches mobile users.
However, the BS requires current channel state information (CSI) to
determine transmit beamforming vectors and user clustering.  Users
with highly correlated channel direction information (CDI) are
assigned to be in the same cluster and share the same transmit
beamforming~\cite{yang17,he21,poor17}.  Assuming perfect CSI, several
existing works~\cite[and references therein]{dhakal19, ding19, xiao18,
  ding17, ali17, wang21, jding20, kim20} studied beamforming design,
power allocation, and user clustering for MIMO-NOMA. However, CSI
errors incurred by quantizing with a finite number of bits or by
estimating the channels can have an adverse effect on the system
performance~\cite{chen17, yang17, Tang20, zhang20, cui18, hoang21,
  ojcoms20}.

For power-domain MIMO-NOMA, recent works focused on imperfect CSI or
finite CSI quantization rate in conjunction with transmit power
allocation~\cite{cui18, ojcoms20, zhang20, chen17, Tang20, nandan21,
  wan21, liuxia18, kota21, gong19}.  Several studies~\cite{zhang20,
  chen17, Tang20, kota21, gong19, gao21, liuxia18} proposed the
transmission schemes that maximize the sum rate of all
users. In~\cite{zhang20}, power allocation for users sharing the same
transmit beam, was optimized for given imperfect CSI.  Both power and
CSI feedback allocations were optimized in~\cite{chen17, Tang20}.  An
imperfect CSI robust beamforming design and power allocation for
millimeter wave MIMO-NOMA was proposed in~\cite{kota21}.
In~\cite{gong19}, power allocation and user clustering were proposed.
A beamforming design with power allocation was studied in~\cite{gao21}
for satellite transmission. A low complexity power allocation scheme
that maximizes the sum rate while ensuring user's fairness was
proposed in~\cite{liuxia18}.

Other studies examined different objectives besides maximizing the sum
rate. In our previous work~\cite{ojcoms20}, the allocation of CDI
rates between 2 groups of users was optimized for a given total CDI
rate to attain max-min fairness on either achievable rate or outage
probability.  In~\cite{cui18}, power allocation and beamforming
vectors were jointly optimized under some outage constraints to
maximize the sum of user's utility, which is a function of the
rate. Eigen beamforming technique was proposed in~\cite{nandan21} to
improve the physical layer security for MIMO-NOMA-based cognitive
radio networks where the power allocation was also optimized.
In~\cite{wan21}, a closed-form optimal power allocation was derived
with the objective function being energy efficiency of the system in
bit-per-joule. The work~\cite{zhu22} applied deep reinforcement
learning to allocate power for each vehicular user in MIMO-NOMA
vehicular edge computing system for which the objective was to
minimize long-term power consumption and latency.

For MIMO-NOMA, the number of users sharing a single beamforming vector
can be greater than 2.  However, as the number of users increases, the
complexity of SIC at the receivers increases while the sum rate
decreases~\cite{zeng17}.  Thus, for this work, we assume that 2 users
with highly correlated spatial channels are paired, and only the user
with higher channel power in the pair, quantizes and feeds CDI back to
the BS.  This can reduce the total CDI rate over all users and is in
contrast with~\cite{chen17, Tang20} in which all users must quantize
and send back their CDI.  To achieve max-min rate fairness among all
users, the BS must optimize the allocation of CDI rate and transmit
power for users.  To reduce the number of optimizing variables,
grouping of user pairs is proposed for the allocation.  The proposed
user-grouping scheme can support any number of user groups and hence,
generalizes our previous scheme~\cite{ojcoms20} in which only 2 groups
of users were considered and only the CDI rate was optimized.

Given the total CDI rate and total transmit power, we develop a
max-min SINR problem optimizing the CDI rate and transmit power for
each user group, and user grouping.  For given grouping of users, we
can find the optimal CDI rate and transmit power for all users by
solving a geometric program (GP).  To reduce the complexity of finding
solutions, we propose a progressive-filling scheme that gradually
allocates the CDI rate and transmit power for each group of users.
Numerical results show that the progressive filling performs close to
the optimum with much less computational complexity.  To further
increase the system performance, we optimize user grouping by applying
K-means to cluster user pairs with similar performance into the same
group.

For MIMO-OMA, there is no user pairing and thus, all users are
assigned different transmit beamforming. Since all users must send
back CDI to the BS, the total CDI rate for OMA can be higher than that
for NOMA.  However, the receiver in OMA is simpler since SIC at the
receivers is not needed.  Similar to MIMO-NOMA, we develop an
allocation problem that maximizes the minimum SINR.  To accommodate an
arbitrary number of users, regularized zeroforcing is applied to find
transmit beamforming for all users.  For systems with a full
load\footnote{The number of users is equal to the number of transmit
antennas.}, we can find the optimal CDI rate and transmit power for
each user group, and the optimal regularizing constant from GP.
Comparing NOMA and OMA, we find that the NOMA scheme can outperform
the OMA scheme by up to 4 dB when the total CDI rate is small or
moderate.  As the total CDI rate increases, the performance gap
between the 2 multiple access schemes decreases and for a lighter
load, OMA can perform better than NOMA.

Our contribution in this work can be summarized as follows.
\begin{itemize}
  \item We derive asymptotic SINR for all users when the number of BS
    transmit antennas tends to infinity.  The derived SINR is shown to
    approximate the actual SINR well when the system size is
    large. For MIMO-NOMA downlink, we found the optimal CDI rate and
    transmit power for all users that maximize the minimum of the
    approximate SINR.  Optimal solutions can be obtained by GP for
    which there are many efficient and fast solvers.  We propose a
    suboptimal progressive filling scheme that performs close to the
    optimum and can reduce the computation time by almost 3 orders of
    magnitude in some cases. To better the allocation and increase the
    system SINR, we propose to group user pairs with an iterative
    scheme based on K-means over all pairs' SINR.

  \item For MIMO-OMA, we employ regularized zeroforcing instead of
    conventional zeroforcing to accommodate systems with the number of
    users greater than that of transmit antennas. For systems with a
    full load, we can solve for the optimal CDI rate, transmit power,
    and regularizing constant.  For systems with arbitrary load, we
    propose a suboptimal scheme that alternately solves
    subproblems. We compare the performance of OMA and NOMA with the
    number of users larger than the number of BS transmit antennas and
    find that with a higher threshold for user pairing, NOMA can
    outperform OMA in all total-CDI rate regimes.
\end{itemize}

This paper is organized as follows.  Section~\ref{sys_mod} introduces
the channel model, zeroforcing beamforming, CDI quantization, and
transmit power allocation. In Section~\ref{sec:alloc}, we derive the
SINR in a large system limit and develop a joint power CDI-rate
allocation problem that maximizes the minimum large-system SINR. In
Section~\ref{sec:joint_pwr_fb}, we propose the optimal solution and
the suboptimal solutions obtained from progressive-filling
scheme. User grouping, which improves the system performance, is
proposed in Section~\ref{sec:grp}. In Section~\ref{sec:oma}, we
propose the allocation schemes for MIMO-OMA. Numerical results are
presented in Section~\ref{num_re}.  Finally, we conclude this work in
Section~\ref{conclude}.

\section{System Model}
\label{sys_mod}

We consider a discrete-time downlink channel in which a BS with $N_t$
transmit antennas transmits message signals to $K$ single-antenna
users.  For a link from each transmit antenna to a user's antenna, we
assume that the signal propagates through a rich-scattering
environment with no line of sight, and that the signal's delay spread
is much smaller than the symbol period.  Thus, a channel impulse
response of each link can be modeled by a single complex Gaussian
random variable with zero mean and unit variance.  We assume that
transmit antennas at the BS are placed sufficiently far apart that
they are independent from one another.  Hence, the channel gains from
all transmit antennas to a user are independent.

For power-domain NOMA, a user near the cell center should be paired
with another user toward the cell edge with a similar channel
direction.  Thus, both users can share a transmit beamforming, but can
be distinguished in the power domain at a receiver.  For pair $k$, let
$\bh_{k;s}$ denote an $N_t \times 1$ channel vector for the user with
a stronger channel or the cell-center user while $\bh_{k;w}$ denotes
the channel vector for the weaker user or the cell-edge user.  Each
entry in a channel vector corresponds to a channel gain from each
transmit antenna to the user.  The correlation between 2 channels
follows the Gauss-Markov model given by~\cite{dhakal19,zhang20}
\begin{equation}
  \label{hk2}
  \bh_{k;w} = c_{k} (\rho_{k} \bh_{k;s} + \sqrt{1 - \rho_{k}^2}
  \bm{e}_{k})
\end{equation}
where $0 < c_{k} \le 1$ is the degradation factor for pair $k$, $0 \le
\rho_{k} \le 1$ is the correlation coefficient for pair $k$, and
$\bm{e}_{k}$ is an $N_t \times 1$ error vector with independent
complex Gaussian entries with zero mean and unit variance.  Note that
the degradation factor $c_k$ is a ratio between the average channel
power of the cell-edge and the cell-center users of pair $k$.  The
value of $c_k$ close to zero indicates that 2 users are far apart from
each other.  Otherwise, the 2 users are close to each other. The
correlation coefficient $\rho_{k}$ is close to 1 if the channels of
the 2 users are highly correlated.\footnote{Our previous
work~\cite{ojcoms20} strictly assumed a perfect channel alignment
between 2 users for all pairs ($\rho_{k} = 1, \forall k$).} Those
users could lie in the same signal direction and hence, their CDI's
are similar.  For better performance, 2 users should be paired if
$c_k$ is small and $\rho_{k}$ is close to 1. The design of user
pairing is important, but is out of the scope of this work.  Some
users are not in pairs since their channels may not be sufficiently
correlated with others.  These users are referred to as
singletons. Let $\bm{h}_l$ denote an $N_t \times 1$ channel vector for
singleton or unpaired user $l$.

For our proposed NOMA schemes in
Sections~\ref{sec:alloc}--\ref{sec:grp}, we assume $M \le N_t$
transmit beamforming vectors.\footnote{In Section~\ref{sec:oma}, we
apply {\em regularized} zeroforcing beamforming. In that case, the
number of transmit beams $M$ can be greater than $N_t$.} If there
exist $M_1 < M$ singletons, $M_1$ beams are assigned for singletons
and the remaining $M-M_1$ beams are for $M-M_1$ user pairs.  Hence,
the total number of served users denoted by $K = M_1 + 2(M-M_1) = 2M -
M_1$.  To compute zeroforcing beamforming vectors and allocate proper
transmit power, the BS requires current CSI from all singletons and
user pairs.  We assume that singletons and pairs feed back their
channel quality information (CQI) referring to $\|\bh_{l}\|$,
$\|\bh_{k;s}\|$, and $c_{k}$, $\forall l, k$, to the BS {\em
  perfectly}.  For CDI, which requires significantly more quantization
bits, each singleton or pair quantizes and feeds back its normalized
channel vector $\bhb_{l} = \bh_l/\|\bh_l\|$ or $\bhb_{k;s} =
\bh_{k;s}/\|\bh_{k;s}\|$ with a finite number of bits.  Let $b_l$ and
$b_k$ be the number of bits to quantize $\bhb_l$ for singleton $l$ and
$\bhb_{k;s}$ for pair $k$, respectively.  We assume channels are
independent block fading and the block length is sufficiently long
that feeding back CDI is meaningful.

With the quantized CDI, the BS forms the $M \times N_t$ matrix $\bHh$
whose rows are transpose of the quantized channel vectors $\bhh_m$.
Zeroforcing beamforming vector for singleton or pair $m$ is denoted by
$\bw_m$, which is the normalized $m$th column of the $N_t \times M$
matrix given by $\bW = \bHh^{\dag} (\bHh \bHh^{\dag})^{-1}$.  With the
set of transmit beamforming vectors, the BS transmits message symbol
$s_l$ for singleton $l$ and applies superposition coding to convey
symbols $s_{k;s}$ and $s_{k,w}$ for strong and weak users of pair $k$,
respectively.  The transmitted symbol
\begin{equation}
  x = \sqrt{p_l} \sum_{l \in \mathcal{S}} \bw_{l} s_{l} + \sum_{k
    \in \mathcal{U}} \bw_{k} (\sqrt{p_{k;s}} s_{k;s} +
  \sqrt{p_{k;w}} s_{k;w})
\end{equation}
where $s_l$, $s_{k;s}$, and $s_{k;w}$ are zero-mean unit-variance
random variables, $p_l$ is transmit power for singleton $l$ with beam
$\bw_l$, $p_{k;s}$ and $p_{k;w}$ are transmit power for strong and
weak users in pair $k$ sharing beam $\bw_k$, $\mathcal{S}$ and
$\mathcal{U}$ are sets of beam indices for singletons and user pairs,
respectively.  Note that $|\mathcal{S}| = M_1$ and $|\mathcal{U}| = M
- M_1$, and $\mathcal{S} \cap \mathcal{U} = \emptyset$. In any
frequency-time resource block, 2 users in each pair will interfere
with each other in time, frequency, and spatial domains, but will be
distinguished in the power domain.  For transmit power allocation
between 2 users in a pair, we apply the fractional-transmit power
allocation (FTPA) proposed by~\cite{Saito13} as follows
\begin{gather}
  p_{k;s} = \frac{c_{k}^{2 \alpha_{k}}}{1+c_{k}^{2\alpha_{k}}} p_k
  , \label{pk1}\\
  p_{k;w} = \frac{1}{1 + c_{k}^{2\alpha_{k}}}
  p_k, \label{pk2}
\end{gather}
where $p_k = p_{k;s} + p_{k;w}, \forall k \in \mathcal{U}$ is the
total transmit power for beam $\bw_k$ and $\alpha_k \ge 0$ is the
power decay factor for pair $k$.  If $\alpha_{k} = 0$, both users in
pair $k$ are allocated equal transmit power $p_{k;s} = p_{k;w} =
p_{k}/2$.  By increasing the decay factor $\alpha_{k}$, more power is
allocated to the weaker user.  This allocation scheme also enforces
the constraint $p_{k;w} \ge p_{k;s}$ of ensuring stability of SIC at
the receiver for the stronger user.

The instantaneous signal-to-interference plus noise ratio (SINR) for
singleton $l$ is given by
\begin{equation}
  \label{gamma_1L}
\gamma_{l} = \frac{p_l |\bh_{l}^{\dag}\bw_{l}|^2}{\sum_{k \ne l} p_k
  |\bh_{l}^{\dag}\bw_{k}|^2 + \sigma_n^2}
\end{equation}
where $\sigma_n^2$ denotes the power of zero-mean additive white
Gaussian noise.  Since the zeroforcing solutions are computed from
quantized CDI, singleton $l$ will suffer from some residual
interference due to quantization error.  The interference can be
reduced if the CDI quantization error is decreased or the number of
bits to quantize the CDI for singleton $l$ is increased.  For the
stronger user of pair $k$, the signal of the weaker user is decoded
first, and then that signal is reconstructed and subtracted from the
received signal.  With perfect SIC, the stronger user can decode its
signal without any interference from the weaker user. Hence, an
expression for the SINR for the stronger user of pair $k$ is similar
to that for a singleton and is given by
\begin{equation}
  \label{gamma_1K}
  \gamma_{k;s} = \frac{p_{k;s} |\bh_{k;s}^{\dag}\bw_{k}|^2} {\sum_{m
      \neq k} p_{m}|\bh_{k;s}^{\dag}\bw_{m}|^2 + \sigma_n^2} .
\end{equation}
For the weaker user with larger transmit power, its signal is decoded
directly by treating all interfering signals as noise and its SINR is
given by
\begin{equation}
  \label{gamma_2K}
  \gamma_{k;w} = \frac{p_{k;w} |\bh_{k;w}^{\dag}\bw_{k}|^2}{\sum_{m
      \ne k} p_m|\bh_{k;w}^{\dag} \bw_{m}|^2 + p_{k;s}
    |\bh_{k;w}^{\dag} \bw_{k}|^2 + \sigma_n^2}.
\end{equation}
From~\eqref{gamma_2K}, we see the additional interference in the
denominator, which accounts for intra-pair interference from the
stronger user in the pair.

We would like to maximize the minimum achievable rate for all users in
the cell to achieve max-min fairness.  The objective can be obtained
by optimizing the transmit power and CDI-quantization bit allocation
with a total transmit-power constraint given by
\begin{equation}
  \sum_{l \in \mathcal{S}} p_l + \sum_{k \in \mathcal{U}} p_k = \Ptot
  \label{eq:ptot}
\end{equation}
where $\Ptot$ is the total transmit power, and a total
CDI-quantization bit constraint is given by
\begin{equation}
  \sum_{l \in \mathcal{S}} b_l + \sum_{k \in \mathcal{U}} b_k =
  \Btot
  \label{eq:btot}
\end{equation}
where $\Btot$ is the total number of quantization bits for CDI or the
number of CDI feedback bits from users to the BS.
From~\eqref{eq:ptot} and~\eqref{eq:btot}, there are $2M$ optimizing
variables.  Optimizing the power decay factor $\alpha_k$
in~\eqref{pk1} and~\eqref{pk2} will add another $M-M_1$ variables.
For systems with a larger number of BS transmit antennas, the number
of optimizing variables totaling $3M - M_1$ can be large. To lessen
the computational complexity, we propose to approximate the problem
and present suboptimal solutions that perform well in the subsequent
sections.

\setcounter{MYtempeqncnt}{\value{equation}}

\begin{figure*}[!b]

\normalsize

\setcounter{equation}{14}

\vspace*{4pt}

\hrulefill

\begin{equation}
  \label{gamma_LargeK2}
  \gamma_{k;w} \to \gamma_{k;w}^{\infty} = \frac{\zeta_g (1-\brM) (1 -
    2^{-\brB_g})}{\brK (1 + c_k^{2\alpha_k})(c_{k}^2 - c_k^2
    \rho_{k}^2(1 - 2^{-\brB_g})) + \frac{\sigma_n^2}{P_{\text{tot}}})
    +\zeta_g c_{k}^{4\alpha_{k}} (1-\brM)(1-2^{-\brB_g})}
\end{equation}

\end{figure*}

\setcounter{equation}{\value{MYtempeqncnt}}

\section{Approximate Allocation Problems}
\label{sec:alloc}

To reduce the number of variables to be optimized, we group all
singletons into the first user group since their SINR
expressions~\eqref{gamma_1L} are similar.  We assign all singletons
with the same transmit power $p_l = P_1$ and the same number of CDI
quantization bits $b_l = B_1$, $\forall l \in \mathcal{S}$.  For a
user pair, its performance will be dominated by the SINR of the weaker
user in~\eqref{gamma_2K}, which largely depends on the degradation
factor $c_k$ and the channel correlation coefficient $\rho_k$.  We
will later in Section~\ref{sec:grp} propose to group user pairs by
theirs SINR.  Assume that there are $G$ groups of users, where the
first group is for all singletons and the other $G-1$ groups are for
user pairs, and $2 \le G \le M$.  Let $\mathcal{G}(\cdot)$ be a
grouping function whose input is either an index of a pair or that of
a singleton and output is a group index $g$, $1 \le g \le G$.  We
allocate the same resource to all pairs in the same group.  Thus, user
pair $k$ in group $g = \mathcal{G}(k)$, will be allocated with
transmit power $p_k = P_g$ and a number of bits $b_k = B_g$.  After
pair grouping, group $g$ is assumed to consist of $M_g$ user pairs.
Hence, $\sum_{g=1}^G M_g = M$ where $M_1$ is the number of singletons.

For the min-max rate fairness, we must allocate transmit power and CDI
bits based on the ergodic rates for all 3 types of users, which depend
on the distribution functions of the SINR in~\eqref{gamma_1L},
\eqref{gamma_1K}, and~\eqref{gamma_2K}.  Due to the intractability of
the distribution functions of those SINR for any finite $N_t$ and $M$,
we approximate the rate by its large-system limit, which is shown to
be a good approximation when the system size is sufficiently
large~\cite{mimo,dai08,ojcoms20}.  To analyze the rate or the SINR in
a large-system limit, we let $N_t$, $M$, and $K$ tend to infinity with
fixed ratios $\brK = K/N_t$ and $\brM = M/N_t$.  By constraining the
total transmit power $\Ptot$ to be finite, as the number of users $K
\to \infty$, we assume that the transmit power for group $g$ decreases
as follows
\begin{equation}
  P_g K \to \zeta_g \Ptot \quad \text{for } g = 1, 2, \ldots, G,
\end{equation}
where $0 < \zeta_g < 1$ denotes the limiting power fraction of the
total power assigned to each pair of group $g$.  Hence, in a
large-system limit, the total-power constraint~\eqref{eq:ptot}
converges to
\begin{equation}
  \label{zetai}
  \sum_{g=1}^{G} \zeta_g \brM_g = \brK
\end{equation}
where $\brM_g = M_g/N_t$ is the normalized number of user pairs in
group $g$.  For meaningful CDI quantizing, the number of bits $B_g$
must increase with the dimension of the CDI, which is $N_t$.  We
denote the normalized number of CDI quantization bits per transmit
antenna by $\brB_g = B_g/N_t$.  Hence, the total-bit
constraint~\eqref{eq:btot} converges to
\begin{equation}
  \sum_{g=1}^{G} \brB_g \brM_g = \hBtot
\end{equation}
where $\hBtot = \Btot/N_t^2$.  We note that $\Btot$ increases
quadratically with $N_t$ since both $\brB_g$ and $\brM_g$ increase
linearly with $N_t$.

With the above assumptions and the limits derived by~\cite{dai08}, we
can obtain the limiting SINR for all 3 types of users as follows.  For
singleton $l$, as $N_t \to \infty$,
\begin{equation}
  \label{gamma_LargeL}
  \gamma_{l} \to \gamma_{l}^{\infty} =
  \frac{\zeta_1(1-\brM)(1-2^{-\brB_1})}{\brK \left( 2^{-\brB_1} +
    \frac{\sigma_n^2}{\Ptot} \right)} .
\end{equation}
Since this expression holds for all singletons, this is why the BS
should group all singletons together and assign them with the same
transmit power and CDI bits.  The limiting SINR~\eqref{gamma_LargeL}
increases with larger power factor $\zeta_1$ or the normalized bits
$\brB_1$, and with smaller load $\brM$ or $\brK$.  The SINR will
converge to 0 when CDI is not available at the BS ($\brB_1 = 0$), or
the cell is at a full load ($\brM = 1$).

For pair $k$ in group $g$, the stronger user achieves the following
limiting SINR
\begin{equation}
  \label{gamma_LargeK1}
  \gamma_{k;s} \to \gamma_{k;s}^{\infty} = \frac{\zeta_g
    c_{k}^{2\alpha_{k}}(1-\brM)(1 - 2^{-\brB_g})}{\brK (1 +
    c_{k}^{2\alpha_{k}}) \left(2^{-\brB_g} +
    \frac{\sigma_n^2}{P_{\text{tot}}}\right) } .
\end{equation}
Similar to~\eqref{gamma_LargeL}, we apply the limits derived
by~\cite{dai08} to obtain~\eqref{gamma_LargeK1}.  In addition to
$\zeta_g$ and $\brB_g$, the SINR for the stronger user in pair $k$
depends on the intra-pair power allocation via the power decay factor
$\alpha_k$.  Same as~\eqref{gamma_LargeL}, the term $2^{-\brB_g}$ in
the denominator of~\eqref{gamma_LargeK1} accounts for the residual
interference due to CDI quantization error.  As the CDI becomes more
accurate, both the singleton or the stronger user in a pair will
suffer less interference from other users.

For the weaker user in a pair, interference will be relatively larger
due to additional interference from the stronger user and larger
interference from other beams.  The latter is caused by the difference
between the CDI of the weaker user and that of the stronger.  The CDI
difference is dictated by the correlation coefficient $\rho_{k}$
in~\eqref{hk2}, which plays a major role in the performance of the
weaker user.  The limiting SINR of the weaker user in pair $k$ can
also be derived with the limits in~\cite{dai08} and is given
by~\eqref{gamma_LargeK2}.

\addtocounter{equation}{1}

Similar to that for the stronger user, the SINR for the weaker user
increases with a larger CDI rate $\brB_g$.  The SINR also increases
with better channel alignment between 2 users in a pair (a larger
$\rho_k$).  We note that the last term in the denominator
in~\eqref{gamma_LargeK2} can be attributed to the interference from
the stronger user's signal, which can be reduced by a smaller
degradation factor $c_k$.  Hence, the performance of a pair that
shares the same beam can be increased by pairing users with a large
channel correlation coefficient ($\rho_k$ close to 1) and a
significant difference in CQI ($c_k$ close to 0).

In a large-system limit, an achievable rate for each user converges to
$\log_2(1 + \gamma^{\infty})$ where $\gamma^{\infty}$ is obtained from
either~\eqref{gamma_LargeL}, \eqref{gamma_LargeK1},
or~\eqref{gamma_LargeK2}.  We would like to maximize the minimum rate
of all users in the cell given the total CDI-rate constraint and the
total transmit-power constraint.  Since the rate increases
monotonically with SINR, we optimize SINR directly over power
fractions $\{\zeta_g\}$, normalized CDI bits $\{ \brB_g \}$, power
decay factors $\{\alpha_k\}$, and grouping function $\mathcal{G}$.
The optimization problem is stated as follows.
\begin{subequations}
  \begin{alignat}{3}
  \max_{\{\zeta_{g}\}, \{\brB_g\}, \{\alpha_{k}\}, \mathcal{G}} &
  \quad && \min_{l \in \mathcal{S}, k \in \mathcal{U}}
  \{\gamma^{\infty}_l, \gamma^{\infty}_{k;s}, \gamma^{\infty}_{k;w} \} && \\
  \text{subject to} &&& \sum_{g=1}^{G} \zeta_g \brM_g \le \brK, &&
  \label{const:pwr}\\
  &&& \sum_{g=1}^{G}\brM_g\brB_g \le \hBtot, &&\label{const:bit}\\
  &&& \zeta_{g} \ge 0, \brB_g \ge 0, \quad g = 1,2,\ldots,G,&&
  \label{zb}\\
  &&& \alpha_k \ge 0, \quad \forall k \in \mathcal{U},&& \label{ak}\\
  &&& \mathcal{G}(m) = \left\{
  \begin{array}{lr}
    1 & : m \in \mathcal{S},\\
    2,3,\ldots, G & : m \in \mathcal{U} .
  \end{array} \right.&&
  \end{alignat}
  \label{eq:opt}
\end{subequations}

This generalizes the problem considered in our previous
work~\cite{ojcoms20} in which there were only 2 groups, i.e., a group
of all singletons and a group of all pairs, and only $\brB_1$ and
$\brB_2$ were optimized.  Also, the work in~\cite{ojcoms20} only
assumed perfect channel alignment between the 2 users in any pair or
$\rho_{k} = 1, \forall k \in \mathcal{U}$.  Finding optimal solutions
for the clustering nonlinear problem in~\eqref{eq:opt} is exceedingly
complex.  Hence, we will propose a suboptimal solution by dividing the
problem into 2 subproblems.  The first subproblem is to optimize the
transmit power and the CDI rate with a fixed pair grouping and the
second subproblem is to optimize the grouping of user pairs with fixed
transmit power and CDI rate.  We will consider the first subproblem in
Section~\ref{sec:joint_pwr_fb} and the second one in
Section~\ref{sec:grp}.

\section{Joint Transmit Power and CDI-Rate Optimization}
\label{sec:joint_pwr_fb}

In this section, we assume that the group assignment for all user
pairs is fixed or the grouping function $\mathcal{G}$ is provided.
Hence, problem~\eqref{eq:opt} is simplified and transformed as
follows.
\begin{subequations}
  \begin{alignat}{3}
  \max_{\{\zeta_{g}\}, \{\brB_g\}, \{\alpha_{k}\}} &
  \quad && \gth && \\
  \text{subject to} &&& \eqref{const:pwr}, \eqref{const:bit},
  \eqref{zb}, \eqref{ak}, && \nonumber\\
  &&& \gamma_l^\infty \ge \gth, &\quad & \forall l \in \mathcal{S},
  \label{opt:l}\\
  &&& \gamma_{k;s}^\infty \ge \gth, \quad \gamma_{k;w}^\infty \ge \gth,
  &\quad & \forall k \in \mathcal{U}. \label{opt:k}
  \end{alignat}
  \label{eq:opt_pwr_cdi}
\end{subequations}
Note that the objective function for problem~\eqref{eq:opt_pwr_cdi} is
the minimum attainable SINR for all users denoted by $\gth$.  With the
additional SINR constraints~\eqref{opt:l} and~\eqref{opt:k}, the
objective functions in~\eqref{eq:opt_pwr_cdi} and~\eqref{eq:opt} are
equivalent.

\subsection{Geometric Program}
\label{sec:gp}
\setcounter{MYtempeqncnt}{\value{equation}}

\begin{figure*}[!b]

\normalsize

\setcounter{equation}{27}

\vspace*{4pt}

\hrulefill

\begin{equation}
  \label{g_kmin}
  \gamma_{k}^\infty = \frac{2 \zeta_g \rho_k^2 ( 1 - \brM)( 1 -
    2^{-\brB_g})} {\sqrt{\brK^2 ( \mathbb{B}_k - 2 \mathbb{C}_k)^2 + 4
      \zeta_g c_{k}^2 \rho_k^2 \brK (1 - \brM)(1 - 2^{-\brB_g})(
      2^{-\brB_g} + \frac{\sigma_n^2}{P_{\text{tot}}}}) +
    \brK(\mathbb{B}_k - 2\mathbb{C}_k)}
\end{equation}

\end{figure*}

\setcounter{equation}{\value{MYtempeqncnt}}

Problem~\eqref{eq:opt_pwr_cdi} can be converted to a geometric program
(GP)~\cite{gpboyd07}, which requires the objective and inequality
constraints to be posynomial.  We substitute~\eqref{gamma_LargeL} into
inequality~\eqref{opt:l} and rearrange the inequality to obtain
\begin{equation}
  \label{initGPL}
  \zeta_1^{-1}\frac{\beta_1}{1-\beta_1} +
  \frac{\sigma_n^2}{P_{\text{tot}}}\zeta_1^{-1}\frac{1}{1-\beta_1}
  \leq Z
\end{equation}
where $\beta_1 \triangleq 2^{-\brB_1}$ and $Z \triangleq
\frac{1-\brM}{\brK \gth}$.  For $\brB_1 > 0$ ($\beta_1 < 1$), we can
expand $\frac{1}{1-\beta_1} = 1 + \sum_{i=1}^{\infty} \beta_1^i$.
With the expansion of $\frac{1}{1-\beta_1}$, \eqref{initGPL} can be
expressed as a posynomial inequality given by
\begin{equation}
  \label{gp_con1}
  \frac{\sigma_n^2}{P_{\text{tot}}} \zeta_1^{-1} Z^{-1} + \left(1 +
  \frac{\sigma_n^2}{P_{\text{tot}}} \right) \zeta_1^{-1} Z^{-1}
  \sum_{i=1}^{\infty} \beta_1^i \leq 1.
\end{equation}

Similarly, we can substitute~\eqref{gamma_LargeK1}
and~\eqref{gamma_LargeK2} into~\eqref{opt:k} and obtain 2 posynomial
constraints on SINR for stronger and weaker users of pair $k$ in group
$g$ given by
\begin{equation}
  \label{gp_con2}
  \zeta_g^{-1} Z^{-1} (1 + a_{k}^{-1})\left[
    \frac{\sigma_n^2}{P_{\text{tot}}} + \left(1 +
    \frac{\sigma_n^2}{P_{\text{tot}}} \right)
    \sum_{i=1}^{\infty}\beta_g^i \right] \leq 1
\end{equation}
where $\beta_g \triangleq 2^{-\brB_g}$ and $a_k \triangleq c_k^{2
  \alpha_k}$, and
\begin{multline}
  \label{gp_con3}
  \left(\frac{1-\brM}{\brK}\right) c_{k}^2 a_{k} Z^{-1} \\
  +
  \left(c_{k}^2 (1 - \rho_{k}^2) + \frac{\sigma_n^2}{P_{\text{tot}}}
  \right) \zeta_g^{-1} Z^{-1} (1 + a_k) \\
  + \left(c_{k}^2 +
  \frac{\sigma_n^2}{P_{\text{tot}}} \right) \zeta_g^{-1} Z^{-1} (1 +
  a_k) \sum_{i=1}^{\infty}\beta_g^i \leq 1,
\end{multline}
respectively.  The constraints on total transmit
power~\eqref{const:pwr} and total CDI rate~\eqref{const:bit} can also
be converted to posynomial inequalities given by
\begin{gather}
  \sum_{g=1}^{G}\frac{\brM_g}{\brK}\zeta_g \leq 1,\label{gp_con4}\\
  2^{-\hBtot}\prod_{g=1}^{G}\beta_g^{-\brM_g} \leq 1. \label{gp_con5}
\end{gather}
With the equivalent posynomial constraints, we can rewrite
problem~\eqref{eq:opt_pwr_cdi} as
\begin{subequations}
  \begin{alignat}{3}
  \min_{\{\zeta_{g}\}, \{a_{k}\},\{\beta_g\}, Z} &
  \quad && Z && \\
  \text{subject to} &&& \eqref{gp_con1},\eqref{gp_con4}, \eqref{gp_con5},
  \nonumber &&\\
  &&& \eqref{gp_con2},\eqref{gp_con3}, &\quad &
  \forall k \in \mathcal{U},
  \nonumber \\
  &&& 0 < a_{k} \leq 1, &\quad & \forall k \in \mathcal{U},\\
  &&& \zeta_{g} > 0, \quad 0 < \beta_g < 1, &\quad &g = 1,2,\ldots,G, \\
  &&& Z > 0 . &&
  \end{alignat}
  \label{joint_opt_gp}
\end{subequations}
Problem~\eqref{joint_opt_gp} is not convex due to its posynomial
constraints.  However, it is a GP and can be solved efficiently by
many readily available software packages, such as {\tt MOSEK}, {\tt
  CVXOPT}, and {\tt GGPLAB}.  One efficient method to solve a GP is to
convert it to a convex problem by logarithmic transformations of
variables and constraint functions~\cite{gpboyd07}.  We use {\tt
  MOSEK} to solve~\eqref{joint_opt_gp} and will show the results in
Section~\ref{num_re}.  Let us denote the optimal solutions
for~\eqref{joint_opt_gp} by $\{\zeta_{g}^*\}$, $\{a_{k}^*\}$,
$\{\beta_g^*\}$, and $Z^*$.  Hence, the optimal normalized CDI
quantization rate is $\brB_g^* = -\log_2(\beta_g^*)$ and the optimal
power decay factor for pair $k$ is $\alpha_{k}^* = -
\ln(a_{k}^*)/\ln(c_{k}^2)$, and the minimum SINR for all users is
$\gth^* = (1-\brM)/(Z^* \brK)$.

Since explicit complexity in polynomial form does not exist for
solving GP~\cite{chassein14}, we determine the complexity of
problem~\eqref{joint_opt_gp} by the number of optimizing variables and
constraints.  There are $2G + M - M_1 + 1$ variables, $2K + 3$
inequality constraints, and another $2G + M - M_1 + 1$ boundary
constraints on the variables.  The complexity of
solving~\eqref{joint_opt_gp} increases with the system size and the
number of user groups.  To compare the computational complexity of
different schemes, we will compare the computation time in
Section~\ref{num_re}.

\subsection{Progressive-Filling Allocation}
\label{sub_joint_opt}

To further reduce the complexity of finding the optimal resource
allocation, we propose to first optimize the power decay factor
$\alpha_k$ for all user pairs for a given power factor $\zeta_g$ and
normalized CDI rate $\brB_g$, and then allocate $\zeta_g$ and $\brB_g$
for all user groups by a progressive-filling scheme.  Progressive
filling has been previously applied to allocate limited resource to
nodes in a network to achieve max-min fairness~\cite{mollah18}.

For pair $k$, we first find the power decay factor $\alpha_k \ge 0$
that maximizes the SINR minimum of the 2 users, $\min \{
\gamma^\infty_{k;s}, \gamma^\infty_{k;w}\}$.

\begin{prop}
\label{prop_optpwr}
For a given $\zeta_g$ and $\brB_g$, if
\begin{multline}
  2^{-\brB_g} \left( 1 - c_{k}^2 + \frac{\zeta_g c_k^2 (1-\brM)}{2
    \rho_{k}^2 \brK} \right)\\
  \ge (\rho_{k}^{-2} - 1) ( c_{k}^2 +
  \frac{\sigma_n^2}{P_{\text{tot}}}) + \frac{\zeta_g c_{k}^2(1 -
    \brM)}{2 \rho_{k}^2 \brK},
  \label{ieq:alp0}
\end{multline}
the optimal power decay factor that maximizes the minimum SINR between
the stronger and weaker users of pair $k$ is $\alpha_{k}^* = 0$ and
the minimum SINR for pair $k$ is given by
\begin{equation}
 \gamma_{k}^\infty = \frac{\zeta_g(1-\brM)(1 - 2^{-\brB_g}) }{2 \brK
   \left( 2^{-\brB_g} + \frac{\sigma_n^2}{P_{\text{tot}}} \right)}.
 \label{eq_gammaku}
\end{equation}

Otherwise, the optimal power decay factor is
\begin{equation}
  \label{opt_decay}
  \alpha_{k}^* = \frac{\ln \left(\sqrt{\mathbb{B}_k^2 - 4\mathbb{A}_k
      \mathbb{C}_k} - \mathbb{B}_k \right) - \ln( 2
    \mathbb{A}_k)}{\ln(c_{k}^2)}
\end{equation}
and the resulting minimum SINR for pair $k$ is given by~\eqref{g_kmin}
where \addtocounter{equation}{1}
\begin{align}
  \mathbb{A}_k &= c_{k}^2 + c_{k}^2 \left( \frac{\zeta_g(1-\brM)}{\brK} -
  \rho_k^2 \right)(1 - 2^{-\brB_g}) +
  \frac{\sigma_n^2}{P_{\text{tot}}}, \label{decay_a}\\
  \mathbb{B}_k &=
  (1-\rho_k^2) \left( c_k^2 + \frac{\sigma_n^2}{P_{\text{tot}}}
  \right) - \rho_{k}^2 (1 - c_{k}^2)
  2^{-\brB_g}, \label{decay_b}\\
  \mathbb{C}_k &= -\rho_{k}^2 \left(
  2^{-\brB_g} + \frac{\sigma_n^2}{P_{\text{tot}}}
  \right). \label{decay_c}
\end{align}
\end{prop}
A proof of the proposition is in Appendix~\ref{proof_prop}.

From the above proposition, if the allocated CDI rate per user pair
$\brB_g$ is so low that \eqref{ieq:alp0} is true, the optimal power
allocation among the 2 users is uniform ($\alpha^*_{k} = 0$).  The
resulting SINR for the stronger user will be smaller than that for the
weaker user, and will dictate the performance of the pair.  If
$\brB_g$ is sufficiently large such that \eqref{ieq:alp0} is false,
the optimal decay factor is given by~\eqref{opt_decay} and both users
achieve the same SINR given by~\eqref{g_kmin}.  Also,
\eqref{opt_decay} is likely false for user pairs with good channel
alignment ($\rho_k^2 \approx 1$) or with a large CQI gap ($c_k^2
\approx 0$).  For user pairs with those properties, the optimal power
allocation will most likely be nonuniform ($\alpha^*_{k} > 0$).

Before progressively allocating the transmit power and CDI bits for
each group, we divide the total power and total bits into small
chunks. Let $N_{\zeta}^{(0)}$ and $N_{B}^{(0)}$ be the initial number
of chunks for power and CDI-bit allocations, respectively.  Both
$N_{\zeta}^{(0)}$ and $N_{B}^{(0)}$ must be greater than the number of
user groups $G$.  First, each group will be allocated with the same
cumulative power factor of $\brK/N_{\zeta}^{(0)}$ and the same
cumulative normalized bits of $\hBtot/N_{B}^{(0)}$.  For group $g$,
the power factor per beam and normalized bits per beam are set to
$\zeta_g = \brK/(N_{\zeta}^{(0)} \brM_g)$ and $\brB_g =
\hBtot/(N_{B}^{(0)} \brM_g)$, respectively.

With this initial allocation, we compute the SINR for a
singleton~\eqref{gamma_LargeL} and the SINR for pair $k$ from
Proposition~\ref{prop_optpwr}. The group with the minimum SINR will be
allocated an additional power factor of $\brK/(N_{\zeta}^{(0)}
\brM_g)$.  We then recalculate the SINR of the group with the
additional power factor and find the group with the minimum SINR,
which will be assigned an additional chunk of CDI bits.  Unlike the
power factor, we propose to allocate the remaining $\hBtot (1 -
G/N_B^{(0)})$ bits as a geometric sequence of $N_B^{(0)} - G$ chunks
with decay factor $0 < \delta < 1$.  Therefore, the chunk size of the
normalized bits in the initial iterations is much larger than that in
the final iterations.  This is due to the small rate of change of the
SINR for each user with respect to the normalized bits $\brB_g$ when
$\brB_g$ is close to zero.  With too small CDI-bit allocation, the
SINR increase will be tiny and the algorithm will take a larger number
of iterations to converge.  Thus, we allocate larger chunks first
followed by smaller chunks.  With a sum of the geometric sequence, we
can derive the chunk size of the normalized CDI bits for the $i$th
iteration given by
\begin{equation}
  \Delta \brB_{i} = \hBtot \left( 1 - \frac{G}{N_B^{(0)}} \right)
  \frac{( 1 - \delta)\delta^{i-1}}{1 - \delta^{N_B^{(0)} - G}}
  \label{eq:bitchunk}
\end{equation}
where $i = 1,2, \ldots, N_B^{(0)} - G$.  After allocating additional
CDI bits, the SINR for the group is re-computed.  We then find the
group with the minimum SINR to allocate the additional power factor.
Power factors and normalized bits are alternately allocated to the
group with the minimum SINR.  In other words, user groups are
progressively filled with both resources until both are exhausted.
The allocation at the end is the solution.  We summarize the scheme in
Algorithm~\ref{al:prog_fill}.

\begin{algorithm}
  \caption{The proposed progressive-filling power and CDI bit
    allocation.}
  \label{al:prog_fill}
\begin{algorithmic}[1]
  \STATE Set $N_{\zeta}^{(0)} > G$, $N_{B}^{(0)} > G$, $0< \delta <
  1$, and $i=0$.

  \STATE Initialize $\brB_g = \frac{\hBtot}{N_{B}^{(0)} \brM_g}$ and
  $\zeta_g = \frac{\brK}{N_{\zeta}^{(0)} \brM_g}$ for $g =
  1,2,\ldots,G$.

  \STATE Compute $\gamma_l^\infty(\zeta_1,\brB_1), \forall l \in
  \mathcal{S}$ from~\eqref{gamma_LargeL} and
  $\gamma_k^\infty(\zeta_g,\brB_g), \forall k \in \mathcal{U}$ from
  Proposition~\ref{prop_optpwr} where $g = \mathcal{G}(k)$.

  \STATE $N_{\zeta} = N_{\zeta}^{(0)} - G$

  \STATE $N_B = N_{B}^{(0)} - G$

  \STATE Find $g^* = \mathcal{G}(m^*)$ where $m^* = \arg \min_{m \in
    \mathcal{S} \cup \mathcal{U}} \gamma_m^\infty$.

  \WHILE{$N_{\zeta}>0$ or $N_B>0$}
  \STATE $i \leftarrow i + 1$

  \IF{$N_{\zeta} > 0$}

  \STATE $\zeta_{g^*} \leftarrow \zeta_{g^*} +
  \frac{\brK}{N_{\zeta}^{(0)} \brM_{g^*}}$

  \STATE $N_{\zeta} \leftarrow N_{\zeta}-1$

  \STATE Re-compute $\gamma_{g^*}^\infty (\zeta_{g^*}, \brB_{g^*})$ from
  either~\eqref{gamma_LargeL} or Proposition~\ref{prop_optpwr}.

  \STATE Find $g^+ = \mathcal{G}(m^+)$ where $m^+ = \arg \min_{m \in
    \mathcal{S} \cup \mathcal{U}} \gamma_m^\infty$.

  \ELSE
  \STATE $g^+ \leftarrow g^*$ \ENDIF

  \IF{$N_B > 0$}

  \STATE $\brB_{g^+} \leftarrow \brB_{g^+} + \hBtot \left( \frac{(1 -
    G/N_B^{(0)})( 1 - \delta)}{1 - \delta^{N_B^{(0)} - G}} \right)
  \delta^{i-1}$

  \STATE $N_B \leftarrow N_B-1$

  \STATE Re-compute $\gamma_{g^+}^\infty (\zeta_{g^+}, \brB_{g^+})$
  from either~\eqref{gamma_LargeL} or Proposition~\ref{prop_optpwr}.

  \STATE Find $g^* = \mathcal{G}(m^*)$ where $m^* = \arg \min_{m \in
    \mathcal{S} \cup \mathcal{U}} \gamma_m^\infty$.

  \ELSE
  \STATE $g^* \leftarrow g^+$
  \ENDIF
  \ENDWHILE

  \RETURN $\brB_g$ and $\zeta_g$ for $g = 1, 2, \ldots, G$.
\end{algorithmic}
\end{algorithm}

Generally, $N_\zeta^{(0)}$ and $N_B^{(0)}$ should be sufficiently
large that the achieved minimum SINR is close to the optimum obtained
by GP from Section~\ref{sec:gp}.  However, as we discussed earlier, we
do not set $N_B^{(0)}$ too large to avoid tiny CDI-rate chunks.  From
the numerical results in Section~\ref{num_re}, $N_B^{(0)}$ can be set
to be much smaller than $N_\zeta^{(0)}$.

The complexity of this progressive-filling scheme depends on the
number of iterations, which is $\max\{N_\zeta^{(0)}, N_B^{(0)}\}$, and
the number of SINR computations, which is $N_\zeta^{(0)} + N_B^{(0)} -
G$, and hence, increases with $N_\zeta^{(0)}$ and $N_B^{(0)}$.

\section{Grouping of User Pairs}
\label{sec:grp}

In addition to jointly optimizing the power factor and CDI-bit
allocation in Section~\ref{sec:joint_pwr_fb}, the minimum SINR can be
further increased by optimizing grouping function $\mathcal{G}$.  To
improve max-min fairness, user pairs with similar channel quality
should be grouped together. The optimized grouping can significantly
increase the system performance, especially when the number of groups
is large.  This motivated us to propose a grouping scheme shown in
Algorithm~\ref{user_grouping_method}.  The scheme applies when the
number of groups $G \ge 3$. For $G=2$, we simply group all user pairs
together.

For our proposed scheme, we first cluster all pairs into $G-1$ groups
with K-means, which is a well-known method of vector quantization.
Here, we apply one-dimensional K-means over the SINR of each pair.  In
each subsequent iteration, we move the pair with the worst SINR, which
is also the system SINR, from the current group to the group with a
lower SINR mean.  We expect the pair will perform better in the next
iteration since it is in a group with a lower SINR mean and will
likely be allocated larger power factor and bit allocation.  The
scheme terminates when there is no pair to move or the SINR increase
$\Delta_\gamma$ is less than threshold $\epsilon_\gamma$.  Beyond the
initial grouping by K-means, at the most one pair will be moved at
each iteration.  Thus, the scheme reduces the likelihood that the
grouping oscillates and does not converge.  From the numerical
results, Algorithm~\ref{user_grouping_method} converges within a few
iterations.

The complexity of Algorithm~\ref{user_grouping_method} depends largely
on the joint power and bit allocation scheme used in line~\ref{optpb}
and the number of iterations, which depends on the threshold
$\epsilon_\gamma$.  The computational complexity increases greatly if
GP is applied instead of the progressive filling.

\begin{algorithm}
  \caption{The proposed user-pair grouping}
  \label{user_grouping_method}
  \begin{algorithmic}[1]
    \STATE Set $G \ge 3$.

    \STATE Initialize $\brB_g = (\hBtot - \brM_1\brB_1)/(\brM -
    \brM_1)$, $g = 2, 3,\ldots, G$. \label{initB}

    \STATE Initialize $\zeta_g = (\brK - \brM_1\zeta_1)/(\brM -
    \brM_1)$, $g = 2, 3,\ldots, G$. \label{initZ}

    \STATE Compute $\gamma_k^\infty, \ \forall k \in \mathcal{U}$ with
    Proposition~\ref{prop_optpwr}.

    \STATE Apply K-means to cluster $\{ \gamma_k^\infty\}$ into $G-1$
    groups and obtain the SINR mean for each group. \label{cluster}

    \STATE Label the groups with index $g = 2, 3, \ldots, G$ by
    ascending SINR mean. (The group with label $g=2$ has the smallest
    SINR mean.) \label{labelg}

    \STATE Define function $\mathcal{G}$ with input pair index $k \in
    \mathcal{U}$ and output group index $g \in \{2, 3, \ldots, G\}$,
    based on the clustering in line~\ref{cluster} and labeling in
    line~\ref{labelg}.

    \STATE Set $\gamma_{\text{th-old}} = 0$, $0 < \epsilon_\gamma \ll
    1$, and $\Delta_\gamma = \epsilon_\gamma + 1$.

    \WHILE{$\Delta_\gamma > \epsilon_\gamma$}

    \STATE \label{optpb}Given $\hBtot$, $\brK$, and
    $\mathcal{G}(\cdot)$, solve for $\zeta_g$, and $\brB_g$, $\forall
    g$ by either GP or Algorithm~\ref{al:prog_fill}.

    \STATE Compute $\gamma_l^\infty, \forall l \in \mathcal{S}$ and
    $\gamma_{k}^\infty, \forall k \in \mathcal{U}$.

    \STATE $\gth = \min_{l \in \mathcal{S}, k \in \mathcal{U}} \{
    \gamma_l^\infty, \gamma_k^{\infty}\}$.

    \IF{$\exists k : \gamma_k^{\infty} = \gth$ and $\mathcal{G}(k) = g
      > 2$}

    \STATE Revise function $\mathcal{G}$ such that $\mathcal{G}(k) = g
    - 1$.

    \ENDIF

    \STATE $\Delta_\gamma = \gth - \gamma_{\text{th-old}}$

    \STATE $\gamma_{\text{th-old}} \gets \gth$

    \ENDWHILE

    \RETURN $\mathcal{G}$.
\end{algorithmic}
\end{algorithm}

\section{Orthogonal Multiple Access}
\label{sec:oma}

In this section, we assume that the system does not pair users whose
spatial channels are highly correlated and hence, the system only
consists of singletons.  To accommodate an arbitrary number of users,
$K$, the BS with $N_t$ transmit antennas applies regularized
zeroforcing beamforming to construct for each user its transmit
beamforming vector~\cite{wagner12}.  Assuming flat Rayleigh fading,
the channel vector for user $n$ is given by $c_n \bh_n$ where $\bh_n$
is an $N_t \times 1$ vector whose entries are complex Gaussian with
zero mean and unit variance, $0 < c_n \le 1$ is the degradation
factor, and $n = 1,2,\ldots, K$.  For a cell-center user, $c_n$ is
close to or equal to 1.  For a cell-edge user, $c_n$ is small and
closer to 0.  Since there is no user pairing, the BS randomly selects
$K$ users and hence, the channel vectors corresponding to those users
are independent.

Inspired by the proposed scheme in Section~\ref{sec:grp}, we group
users with similar degradation factor $c_n$ into a group and allocate
resources to each group based on its performance.  We propose that
user $l$ in group $g$ is assigned $B_g$ bits to quantize its CDI,
which will be fed back to the BS.  Similar to NOMA transmission, the
BS forms the $K \times N_t$ quantized channel matrix
$\bHh_{\mathrm{oma}}$ whose $n$th row is a transpose of quantized CDI
for user $n$ denoted by $\bhh_n$.  For the same CDI accuracy, this
scheme will require a much higher total quantization rate than the
NOMA scheme presented in previous sections since the number of CDI
vectors is generally higher.  The BS computes a regularized
zeroforcing beamforming for user $n$ given by $\bv_n = \bV \bhh_n$
where $\bV = (\bHh_{\mathrm{oma}}^\dag \bHh_{\mathrm{oma}} + \varphi
N_t \bm{I} )^{-1}$ and $\varphi$ is a regularizing constant. The
regularized zeroforcing vectors obtained are semi-orthogonal with one
another and hence, induce less interference.  As the CDI becomes more
accurate (as $B_g$ increases), the beamforming vectors become more
orthogonal.  Thus, this is an OMA scheme.

We can derive an expression for the SINR for user $l$ as follows
\begin{equation}
  \label{g_sdma}
  \gamma_n = \frac{p_n c_n^2| \bh_n^{\dag} \bv_n|^2/\|\bv_n\|^2}
        {\sum_{j \neq n} p_j c_n^2|\bh_n^{\dag} \bv_j|^2/\|\bv_j\|^2 +
          \sigma_n^2}
\end{equation}
where $p_n$ is the transmit power for user $n$ and $p_j$ is the
interfering power from user $j$.  If user $n$ belongs in group $g$,
the base station allocates power $p_n = \zeta_g \Ptot/K$ where $\Ptot$
is the total transmit power and $\zeta_g$ is the power factor for
group $g$.  With the results in~\cite{wagner12}, we can show that as
$N_t \to \infty$,
\begin{equation}
  \gamma_n - \gamma_n^\infty \to 0
\end{equation}
where
\begin{equation}
  \label{g_sdma_large2}
  \gamma_{n}^\infty = \frac{\zeta_g (1-2^{-\brB_g}) (\mo)^2 \ao} {1 -
    \mo(\mo + 2)2^{-\brB_g} + (1 + \mo)^2 \sigma_n^2 / (c_n^2
    P_{\text{tot}})},
\end{equation}
\begin{align}
  \mo &= \frac{1}{2} \sqrt{\left(1 + \frac{\brK - 1}{\varphi}
    \right)^2 + \frac{4}{\varphi}} - \frac{1}{2}\left(1 + \frac{\brK -
    1}{\varphi} \right),\label{mo}\\
  \ao & = \left( \sqrt{\brK} +
  \frac{\varphi (1 + \mo)}{\sqrt{\brK}} \right)^2 - 1 .\label{aka}
\end{align}
The asymptotic SINR in~\eqref{g_sdma_large2} clearly increases with
$\zeta_g$, $\brB_g$, and signal-to-noise ratio (SNR) $c^2_n \Ptot /
\sigma^2_n$.  For moderate to large $N_t$, the asymptotic SINR
$\gamma_{n}^\infty$ is shown to be close to the actual SINR $\gamma_n$
\cite{wagner12}. For tractability, we will use the asymptotic SINR to
find the optimal sets of power factors $\{\zeta_g\}$ and the
normalized CDI bits $\{\brB_g\}$.  To achieve max-min fairness for
this OMA scheme, we would like to solve the following problem
\begin{subequations}
  \begin{alignat}{3}
  \max_{\{\zeta_{g}\}, \{\brB_g\}, \varphi, \mathcal{F}} &
  \quad && \min_{n = 1,2,\ldots,K}
  \gamma^{\infty}_n& \\
  \text{subject to} &&& \sum_{g=1}^{G} \zeta_g \brM_g \le \brK, &\\
  &&& \sum_{g=1}^{G}\brM_g\brB_g \le \hBtot, &&\\
  &&& \zeta_{g} \ge 0, \brB_g \ge 0,\quad g = 1,2,\ldots,G,& \\
  &&& \varphi > 0, &\\
  &&& \mathcal{F}: \{1,2,\ldots,K\} \to \{1,2,\ldots,G\}&
  \end{alignat}
  \label{eq:opt_oma}%
\end{subequations}
where $\mathcal{F}$ is a user-grouping function.  Since
$\gamma_{n}^\infty$ monotonically decreases with degradation factor
$c_n$, the minimum SINR for group $g$ is from the user with the
minimum $c^2_n$ in the group denoted by
\begin{equation}
  n^*_g = \arg \min_{n: \mathcal{F}(n) = g} c^2_n .
  \label{eqng}
\end{equation}
This set of users, $\{n^*_g \}$, will determine the resource
allocation for all groups.  In this study, we do not directly solve
problem~\eqref{eq:opt_oma}, but focus on subproblems that optimize
$\{\zeta_g\}$, $\{\brB_g\}$, and $\varphi$, with a given user grouping
$\mathcal{F}$.  The subproblems are discussed next.

\subsection{Optimizing $\zeta_g$, $\brB_g$, and $\varphi$}

The first subproblem of~\eqref{eq:opt_oma} is to jointly optimize
$\zeta_g$ and $\brB_g$ for given regularizing constant $\varphi$ and
grouping function $\mathcal{F}$. This subproblem can also be solved by
GP.  Similar to problem~\eqref{eq:opt_pwr_cdi}, SINR for all users
must exceed a threshold $\gth$. Equivalently, the minimum SINR for
each group $\gamma^\infty_{n_g^*} \ge \gth$.
With~\eqref{g_sdma_large2}, we can rewrite the inequality as follows
\begin{equation}
  \label{ineq_sdma}
  \frac{1 + (1 + \mo)^2\sigma_n^2/(c_{n_g^*}^2 \Ptot)}{\zeta_g(1 -
    2^{-\brB_g}) \ao} + \frac{\mo(\mo + 2) 2^{-\brB_g}}{\zeta_g (1 -
    2^{-\brB_g}) \ao} \leq \frac{(\mo)^2}{\gth} .
\end{equation}
Defining $\beta_g \triangleq 2^{-\brB_g}$ and $Y \triangleq
\frac{(\mo)^2}{\gth}$, we can rewrite~\eqref{ineq_sdma} with
posynomials of $\zeta_g$, $\beta_g$, and $Y$ given by
\begin{multline}
  \label{sdma_gp_con1}
  \left(1 + \frac{(1 + \mo)^2\sigma_n^2}{c_{n^*_g}^2 \Ptot}
  \right)(\ao)^{-1} \zeta_g^{-1} (1 + \sum_{i=1}^{\infty} \beta_g^i)
  Y^{-1} \\ + \mo(\mo+2) (\ao)^{-1} \zeta^{-1} (\sum_{i=1}^{\infty}
  \beta_g^i ) Y^{-1} \leq 1.
\end{multline}
Similar to problem~\eqref{joint_opt_gp}, we can describe this problem
so that the objective and constraints are composed of posynomials of
the variables to be optimized as shown
\begin{subequations}
  \begin{alignat}{3}
  \min_{\{\zeta_{g}\},\{\beta_g\}, Y} &
  \quad && Y && \\
  \text{subject to} &&& \eqref{sdma_gp_con1},
    &\quad &g = 1,2,\ldots,G, \nonumber\\
  &&&\eqref{gp_con4} \text{ and } \eqref{gp_con5}, && \nonumber\\
  &&& \zeta_{g} > 0, \ 0 < \beta_g < 1, &\quad &g = 1,2,\ldots,G, \\
  &&& Y > 0  &&
  \end{alignat}
  \label{joint_opt_gp2}%
\end{subequations}
where~\eqref{gp_con4} and~\eqref{gp_con5} are the constraints on the
total power factor and the total normalized CDI bits,
respectively. Hence, problem~\eqref{joint_opt_gp2} can be solved by a
GP solver.

To further improve the performance, we can optimize regularizing
constant $\varphi$.  The optimal $\varphi$ was derived
from~\cite{wagner12}.  With the optimized $\varphi$, we re-solve
problem~\eqref{joint_opt_gp2} to obtain the new power factor and bit
allocations.  The 2 subproblems are alternately solved until the
minimum SINR converges.  The steps are summarized in
Algorithm~\ref{sub_oma}.  We note that the initial value for $\varphi$
in line~\ref{initvarphi} is the optimal value when CDI is perfect or
$\brB_g \to \infty$ while that in line~\ref{opt_reg} is the optimal
value with quantized CDI~\cite{wagner12}.  Algorithm~\ref{sub_oma}
terminates when the SINR difference from the previous iteration is
less than threshold $\epsilon_\gamma$.

\begin{algorithm}
  \caption{The proposed suboptimal scheme to find $\{\zeta_{g}\}$,
    $\{\brB_{g}\}$, and $\varphi$.}
  \label{sub_oma}
  \begin{algorithmic}[1]
    \STATE Initialize $\varphi = \frac{\brK
      \sigma_n^2}{\Ptot}$. \label{initvarphi}

    \STATE Set $\gamma_{\text{th-old}} = 0$, $0 < \epsilon_\gamma \ll
    1$, and $\Delta_\gamma = \epsilon_\gamma + 1$.

    \WHILE{$\Delta_\gamma > \epsilon_\gamma$}

    \STATE Solve~\eqref{joint_opt_gp2} to obtain $\{\zeta_g\}$ and
    $\{\brB_g\}$.

    \STATE Compute $\gamma_{n_g^*}^\infty$ for $g = 1, 2, \ldots, G$
    where $n_g^*$ is obtained by~\eqref{eqng}.

    \STATE $\displaystyle \gth = \min_{g=1,2,\dots,G} \gamma_{n_g^*}^\infty$ and
    $\displaystyle g_{\min} = \arg \min_{g=1,2,\dots,G} \gamma_{n_g^*}^\infty$

    \STATE $\varphi = \frac{c^2_{n_{g_{\min}}^*} \Ptot/\sigma^2_n +
      2^{-\brB_{g_{\min}}}}{1 - 2^{-\brB_{g_{\min}}}}$. \label{opt_reg}

    \STATE $\Delta_\gamma = |\gth - \gamma_{\text{th-old}}|$.

    \ENDWHILE

    \RETURN $\{\zeta_{g}\}$, $\{\brB_{g}\}$, and $\varphi$.
\end{algorithmic}
\end{algorithm}

\subsection{Full Load ($\brK = 1$)}

Next, we consider a system with a full load or with the number of
users equal to the number of transmit antennas at the BS.  Substitute
$\brK = 1$ into~\eqref{mo} and~\eqref{aka} to obtain
\begin{align}
  \mo &= \frac{1}{2} \sqrt{1 + \frac{4}{\varphi}} - \frac{1}{2},
    \label{mo1}\\
  \ao & = \frac{2\mo + 1}{(\mo)^2}. \label{ao1}
\end{align}
We let
\begin{align}
  d &= 2 \mo + 1 = \sqrt{1 + \frac{4}{\varphi}}, \label{d2}\\
  S &= \frac{1}{\gth}. \label{s2}
\end{align}
Substituting~\eqref{mo1}-\eqref{s2} into~\eqref{ineq_sdma}, we can
express inequality~\eqref{ineq_sdma} in terms of posynomials of
$\zeta_g$, $\beta_g$, $d$, and $S$ as follows
\begin{multline}
  \label{sdma_gp_con2}
  \zeta_g^{-1} {S}^{-1} (2 + d +
  d^{-1}) \left(\sum_{i=1}^{\infty} \beta_g^i + \frac{(1 +
    \sum_{i=1}^{\infty} \beta_g^i) \sigma_n^2}{4 c_{n_g^*}^{2}
    \Ptot}\right)\\ + \zeta_g^{-1} {S}^{-1} d^{-1}\leq 1.
\end{multline}
For $\brK = 1$, we can describe a problem that solves for max-min
fairness as follows :
\begin{subequations}
  \begin{alignat}{3}
  \min_{\{\zeta_{g}\},\{\beta_g\}, d, S} &
  \quad && S && \\
  \text{subject to} &&& \eqref{sdma_gp_con2},
    &\quad &g = 1,2,\ldots,G, \nonumber\\
  &&&\eqref{gp_con4} \text{ and } \eqref{gp_con5}, && \nonumber\\
  &&& \zeta_{g} > 0, \ 0 < \beta_g < 1, &\quad &g = 1,2,\ldots,G, \\
  &&& S > 0, \ d > 1 .&&
  \end{alignat}
  \label{opt_sdma_gp2}%
\end{subequations}
After solving~\eqref{opt_sdma_gp2} with GP, the optimized regularizing
constant $\varphi^* = 4/((d^*)^2 - 1)$ is obtained along with the
optimized power factors and bit allocations.  Thus, $\varphi$ does not
need to be solved separately.

\section{Numerical Results}
\label{num_re}

\begin{figure}[ht]
\centering
\includegraphics[width=\myfigwidth]{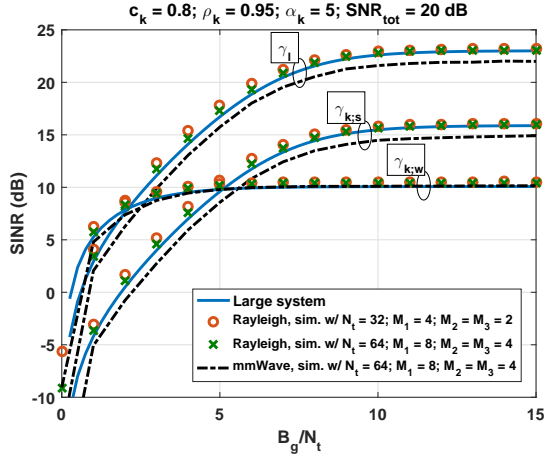}
  \caption{SINR for singleton, $\gamma_{l}$, and that for the stronger
  user in pair $k$, $\gamma_{k;s}$ and the weaker user, $\gamma_{k;w}$
   for finite-size systems are compared with the large-system results.}
\label{f1-SINR_3}
\end{figure}

In Fig.~\ref{f1-SINR_3}, we compare the large-system SINR for a
singleton~\eqref{gamma_LargeL}, and the stronger
user~\eqref{gamma_LargeK1} and the weaker user~\eqref{gamma_LargeK2}
in a pair with the corresponding SINR for finite-size systems obtained
by simulation.  We consider a system with $G=3$, $\brM_1 = 0.125$,
$\brM_2 = \brM_2 = 0.0625$, $\brK = 0.375$, and $\Ptot/\sigma_n^2 =
20$ dB.  The number of transmit antennas at the BS is either $N_t =
32$ or $64$.  For the user pair, degradation factor $c_{k} = 0.8$,
channel correlation coefficient $\rho_{k} = 0.95$, and the decay
factor $\alpha_k = 5$.  As expected, SINR for all 3 users increases
with the normalized CDI bit rate per antenna ($\brB_1$ for a singleton
and $\brB_2$ for a user pair).  We see that a singleton achieves
larger SINR $\gamma_{l}$ when the CDI rate is high since it does not
share the transmit beam with other users and therefore, does not
suffer additional interference.  With a large power decay factor
$\alpha_k =5$, more power is allocated for the weak user in the pair
and hence, in the small CDI-rate regimes, SINR for the weak user is
larger than that for the strong user.  As the CDI rate increases and
the channel information becomes more accurate, the strong user with
SIC performs better due to more accurate transmit beamforming and less
interference from other users.

As the system size increases from $N_t = 32$ to $64$, the SINR
obtained by simulation approaches the large-system SINR
in~\eqref{gamma_LargeL}, \eqref{gamma_LargeK1}, and
\eqref{gamma_LargeK2}. In addition to the simulation for Rayleigh
fading channels, we also obtain the simulation results for millimeter
wave (mmWave) channels utilized in many mobile networks including
long-term evolution (LTE) networks. For this figure with moderate load
($\bar{K} = 0.375$), SINR for all 3 users in mmWave channels is a bit
smaller than that in Rayleigh fading, but follows the same trend as
the CDI rate increases.  The model for mmWave channels in this work is
from~\cite{kota21} with 20 signal paths and an exponential power-delay
profile. From our numerical study, we find that for a
lighter-to-moderate load or a high CDI rate, our large-system analysis
can reasonably approximate the performance of mmWave channels.  Thus,
the optimized allocation derived in this work is expected to also work
well for systems operated in mmWave bands in those applicable regimes.

\begin{figure}[ht]
\centering
\includegraphics[width=\myfigwidth]{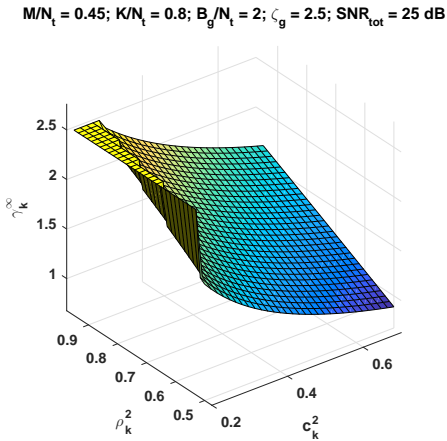}
\caption{A surface plot of $\gamma_{k}^\infty$ obtained from
  Proposition~\ref{prop_optpwr} with varying $c_k^2$ and $\rho_k^2$
  for a system with $\brM=0.45$, $\brK=0.8$, and
  $P_{\text{tot}}/\sigma_n^2 = 25$ dB.}
\label{f12-Surface}
\end{figure}

In Fig.~\ref{f12-Surface}, we show a plot of $\gamma_{k}^\infty$
obtained from Proposition~\ref{prop_optpwr} with varying $c_k^2$ and
$\rho_k^2$ for a system with $\brM=0.45$, $\brK=0.8$, and
$P_{\text{tot}}/\sigma_n^2 = 25$ dB.  If condition~\eqref{ieq:alp0}
applies, the expression for the SINR for pair $k$~\eqref{eq_gammaku}
does not depend on $c_k^2$ and $\rho^2_k$.  Thus, we see a flat SINR
surface.  Otherwise, the SINR expression~\eqref{g_kmin} applies and
the SINR decreases with larger $c^2_k$ and smaller $\rho^2_k$.

\begin{figure}[ht]
\centering
\includegraphics[width=\myfigwidth]{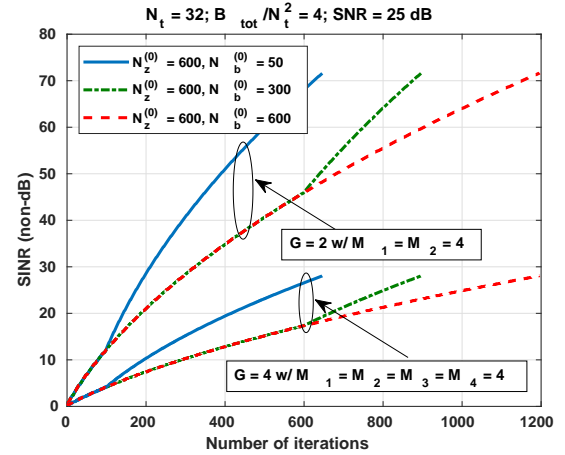}
\caption{The minimum SINR from Algorithm~\ref{al:prog_fill} is plotted
  with the number of iterations for systems with either $G=2$ or
  $G=4$, $N_t=32$, $\hBtot = 4$, $P_{\text{tot}}/\sigma_n^2=25$ dB,
  and $\delta = 0.6$.}
\label{f13-NzNb}
\end{figure}

In Fig.~\ref{f13-NzNb}, we track the minimum SINR obtained from
Algorithm~\ref{al:prog_fill} with the number of iterations for systems
with 2 groups with $M_1 = M_2 = 4$, and 4 groups with $M_1 = M_2 = M_3
= M_3 = 4$.  In the figure, $N_{\zeta}^{(0)} = 500$ while
$N_{B}^{(0)}$ is set to either 50, 300, or 600.  The minimum SINR of
all users increases with the number of iterations.  The maximum of the
minimum SINR is attained when all bit and power chunks are allocated
to the user pairs. The system with 4 groups achieves lower SINR due to
heavier load. From the results shown, setting $N_{B}^{(0)} = 50$
results in the fastest convergence with about 650 iterations.

\begin{figure}[h]
\centering
\includegraphics[width=\myfigwidth]{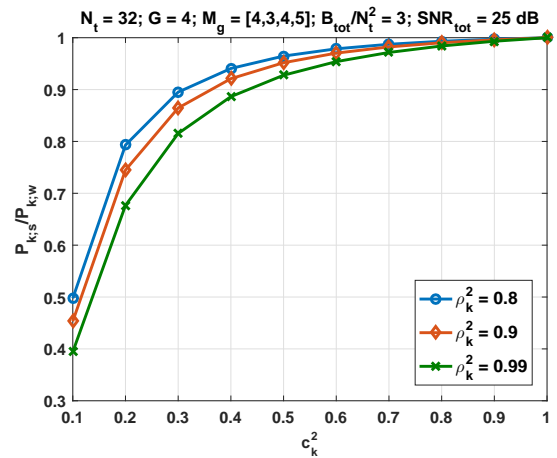}
\caption{For a given pair, ratio $p_{k;s}/p_{k;w}$ is plotted with
  $c^2_k$ for various values of $\rho^2_k$.  $N_t = 32$, $G=4$,
  $\{M_g\} = \{4,3,4,5\}$, $K = 32$, $\hBtot = 3$, and
  $P_{\text{tot}}/\sigma_n^2 = 25$ dB.}
\label{f3-P1toP2}
\end{figure}

For Fig.~\ref{f3-P1toP2}, we solve problem~\eqref{joint_opt_gp} with
GP solver {\tt MOSEK} for a system with 4 user groups and $N_t = 32$,
and obtain the optimal $\{\alpha_{k}^*\}$, $\{\zeta_g^*\}$, and
$\{\brB_g^*\}$.  For one user pair, the ratio between the optimal
power for strong and weak users, $p_{k;s}/p_{k;w}$ is plotted with
varying squared degradation factor $c^2_k$.  As expected, the weak or
the cell-edge user must be allocated with higher power than the strong
or the cell-center user.  As the weak user moves toward the cell
center ($c^2_k$ increases to 1), the power assigned to both users
becomes more uniform.  We also vary the channel correlation between 2
users and note that the weaker user is allocated with a higher power
fraction as $\rho^2_k$ increases.  This is due to stronger intra-pair
interference from the stronger user as both channels become more
aligned.

\begin{figure}[h]
  \begin{subfigure}[t]{0.5\textwidth}
    \centering
    \includegraphics[width=\mysubfigwidth]{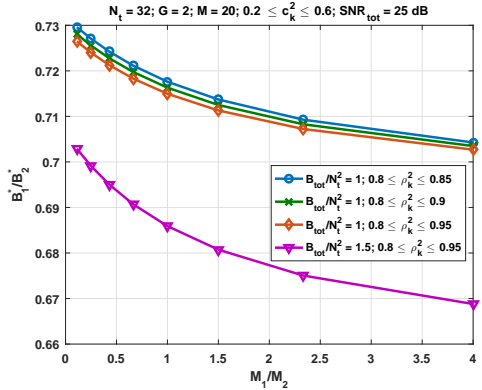}
    \caption{$\brB_1^*/\brB_2^*$ versus $M_1/M_2$}
    \label{f4-OptBit}
  \end{subfigure}
  \hspace{0.5cm}
  \begin{subfigure}[t]{0.5\textwidth}
    \centering
    \includegraphics[width=\mysubfigwidth]{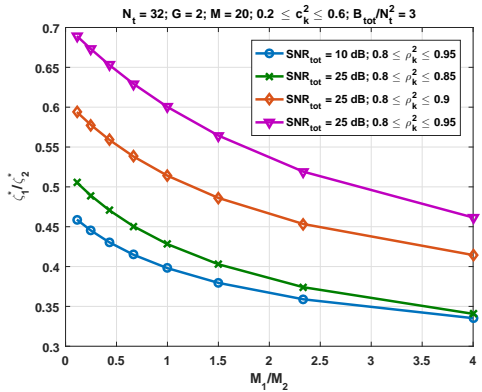}
    \caption{$\zeta_1^*/\zeta_2^*$ versus $M_1/M_2$}
    \label{f5-OptZeta}
  \end{subfigure}
  \caption{$N_t=32$, $G=2$, and $M=20$. For group 2, $0.2 \le c_{k}^2
    \le 0.6$ and $\rho_{k}^2 \ge 0.8$.}
\label{fig_opt_ZF2}
\end{figure}

For Fig.~\ref{fig_opt_ZF2}, we consider a system with 2 groups
($G=2$), $N_t=32$, $M=20$, and vary the number of transmit beams for
group 1 or $M_1$ ($M_2 = M - M_1$). A GP in~\eqref{joint_opt_gp} is
solved.  The ratio between the optimized CDI rates for the 2 groups is
shown with varying $M_1/M_2$ in Fig.~\ref{f4-OptBit} and the ratio
between the optimized power factors is shown in Fig.~\ref{f5-OptZeta}.
Since both ratios are less than 1 for all settings shown, we conclude
that more resources should be allocated to user pairs instead of
singletons.  User pairs must combat the intra-pair interference in
addition to the interference from other users.  From
Fig.~\ref{f4-OptBit}, even larger fraction of the total CDI rate
should be assigned to user pairs when the total CDI rate is high or
when the channel alignment between the 2 users in a pair is better
($\rho^2_k$ close to 1).  Since intra-beam interference is strong when
$\rho^2_k$ is high, a user pair will require a higher CDI rate to
better suppress inter-beam interference. Hence, with fixed $\hBtot =
1$, $\brB_1^*/\brB_2^*$ decreases with increasing $\rho^2_k$.  For
Fig.~\ref{f5-OptZeta}, we fix the normalized total CDI rate $\hBtot =
3$.  When the total SNR is lowered to 10 dB, $\zeta_1^*/\zeta_2^*$ is
smaller.  Therefore, an even higher fraction of the total power should
be devoted to user pairs.  As $M_1/M_2$ increases with fixed $M = 20$,
the total number of users $K$ decreases.  Due to a lighter load, the
interference among users is reduced.  Hence, more resource can be
devoted to user pairs.  As a result, the ratio between the resource
for singleton and that for user pair decreases in both
Figs.~\ref{f4-OptBit} and~\ref{f5-OptZeta}.

\begin{figure}[h]
  \begin{subfigure}[t]{0.5\textwidth}
    \centering
    \includegraphics[width=\mysubfigwidth]{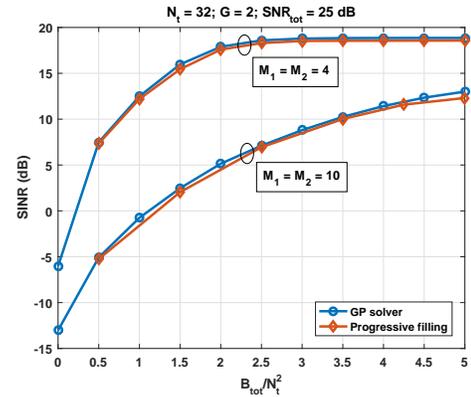}
    \caption{Minimum SINR for the system.}
    \label{f9-G2}
  \end{subfigure}
  \hfill
  \begin{subfigure}[t]{0.5\textwidth}
    \centering
    \includegraphics[width=\mysubfigwidth]{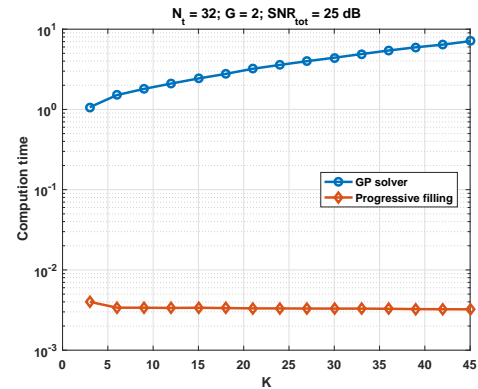}
    \caption{Computation time.}
    \label{f15-Complex}
  \end{subfigure}
  \caption{The SINR comparison between solving ~\eqref{joint_opt_gp2}
    with GP solver {\tt MOSEK} and Algorithm~\ref{al:prog_fill} in
    conjunction with Proposition~\ref{prop_optpwr} for a system with
    $N_t=32$ and $P_{\text{tot}}/\sigma_n^2 = 25$ dB.}
\label{fig_opt2}
\end{figure}

In Fig.~\ref{fig_opt2}, we compare the performance of 2 allocation
schemes from Section~\ref{sec:joint_pwr_fb}.  The GP solutions are
shown with blue lines and circular markers and solutions from the
progressive-filling scheme with Algorithm~\ref{al:prog_fill} in
conjunction with Proposition~\ref{prop_optpwr} are shown with orange
lines and diamond markers.  In Fig~\ref{f9-G2}, the system SINR
increases with a larger total CDI rate and smaller load as
expected. For a lighter load $M = M_1+M_2 =8$, the normalized total
CDI rate $\hBtot$ of approximately 2 is sufficient to achieve close to
the maximum performance.  For a larger load, much higher $\hBtot$ is
required.  Although the progressive-filling scheme is suboptimal, it
can perform close to the optimal solutions obtained by GP.  The
computational complexity measured by the computation time is shown in
Fig~\ref{f15-Complex}.  For the same set of parameters, the
progressive filling in Algorithm~\ref{al:prog_fill} is less complex
than {\tt MOSEK}, which is the GP solver employed, by more than 2
orders of magnitude.  As the number of users increases, the
computation time required for {\tt MOSEK} steadily increases while
that for Algorithm~\ref{al:prog_fill} remains approximately constant.

\begin{figure}[h]
\centering
\includegraphics[width=\myfigwidth]{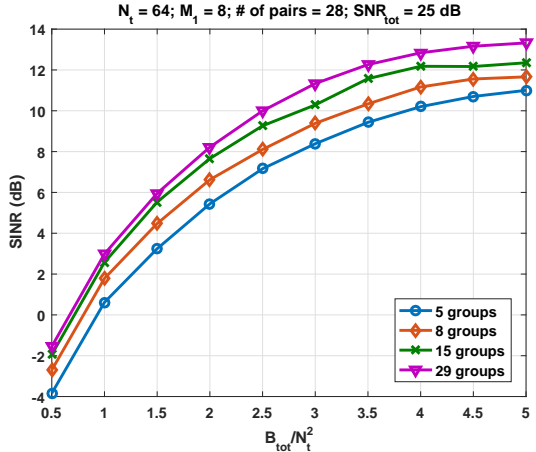}
\caption{The system SINR obtained by
  Algorithm~\ref{user_grouping_method} in conjunction with
  Algorithm~\ref{al:prog_fill} is shown with varying the number of
  groups, $G$.}
\label{f14-NumG}
\end{figure}

Fig.~\ref{f14-NumG} illustrates the performance gain from user
grouping by applying Algorithm~\ref{user_grouping_method} in
conjunction with Algorithm~\ref{al:prog_fill}.  For the figure, the
number of singletons is 8 and that of user pairs is 28 with $N_t =
64$.  We vary the number of groups $G$ from 5 (4 groups of user pairs)
to 29 (each beam gets its own optimized power and CDI rate).  With
increasing $G$, the system SINR increases as expected.  We see a
larger gain when the total CDI rate is moderate ($\hBtot \approx
2.5$). The system gains about 2 dB when $G$ is increased from 5 to 15
and a fraction of a dB when $G$ is increased further to 29.  However,
the complexity of finding these solutions will also increase with $G$.
Thus, there is a trade-off between the SINR and the computational
complexity.

\begin{figure}[ht]
\centering
\includegraphics[width=\myfigwidth]{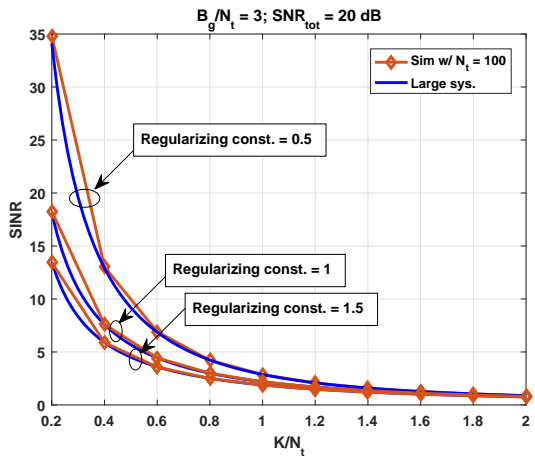}
\caption{The asymptotic SINR~\eqref{g_sdma_large2} is compared with
  the SINR from numerical simulation with $N_t = 100$, $\brB = 3$,
  $P_{\text{tot}}/\sigma_n^2 = 20$ dB, and varying $\varphi$ and
  $\brK$.}
\label{f6-SINR_SDMA}
\end{figure}

In Fig.~\ref{f6-SINR_SDMA}, we compare
the SINR obtained from the large-system analysis~\eqref{g_sdma_large2}
with that from numerical simulation with $N_t=100$, $\brB = 3$, $c_n^2
P_{\text{tot}}/\sigma_n^2 = 20$ dB, and various values of $\varphi$.
As the load $\brK$ increases, SINR decreases, as expected.  For a
lighter load (small $\brK$), SINR also decreases substantially with
larger $\varphi$.  We remark that the large-system result gives an
accurate approximation of the simulation result with large $N_t$.

\begin{figure}[h]
\centering
\includegraphics[width=\myfigwidth]{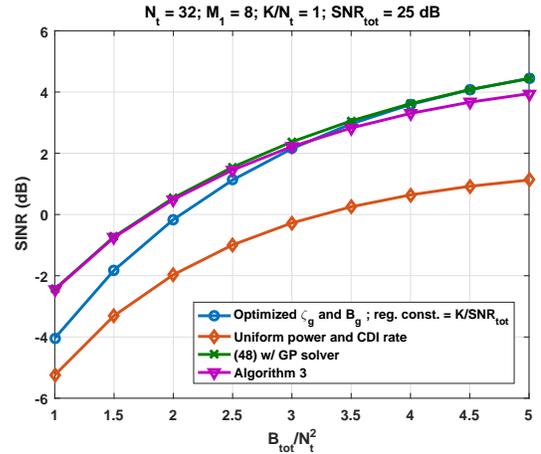}
\caption{The SINR performance with different OMA schemes for a system
  with $N_t= K =32$, $G = 4$ with $M_1 = 8$ and
  $P_{\text{tot}}/\sigma_n^2 = 25$ dB.}
\label{f7-SDMAKbarEq1}
\end{figure}

The performance of OMA schemes described in Section~\ref{sec:oma} is
shown in Fig.~\ref{f7-SDMAKbarEq1}.  For the figure, we set $N_t= K =
32$ ($\brK=1$), and $G = 4$.  Since $\brK = 1$, we can directly solve
for both the allocation $\{\zeta_g\}$ and $\{\brB_g\}$, and
regularizing constant $\varphi$ from problem~\eqref{opt_sdma_gp2} with
a GP solver.  User grouping is based on the degradation factor $c_k$.
The resulting SINR is indicated by the green line in the figure.  We
also apply the suboptimal scheme from Algorithm~\ref{sub_oma} and
obtain the purple curve with triangle markers.  The SINR from
Algorithm~\ref{sub_oma} is very close to the optimum.  Unlike
problem~\eqref{opt_sdma_gp2}, Algorithm~\ref{sub_oma} can be applied
to systems with arbitrary $\brK$.  If $\varphi$ is fixed while
$\{\zeta_g\}$ and $\{\brB_g\}$ are optimized, the performance loss
(the blue curve with circular markers) is noticeable with a larger
total CDI rate $\hBtot$.  Finally, the orange curve with diamond
markers indicates the system SINR with uniform power and CDI rate with
fixed $\varphi = K\sigma_n^2/\Ptot$.  The SINR with uniform allocation
performs the worst across all values of $\hBtot$.

\begin{figure}[h]
  \begin{subfigure}{0.5\textwidth}
    \centering
    \includegraphics[width=\mysubfigwidth]{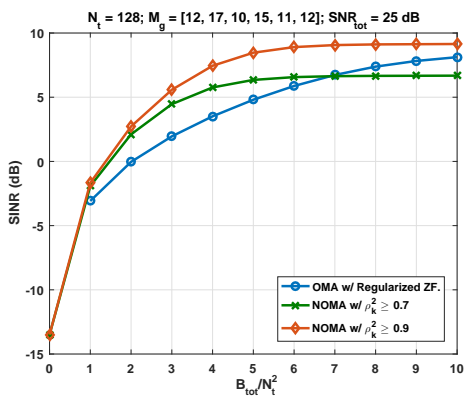}
    \caption{$K = 142$}
    \label{f10-G61}
  \end{subfigure}
  \hfill
  \begin{subfigure}{0.5\textwidth}
    \centering
    \includegraphics[width=\mysubfigwidth]{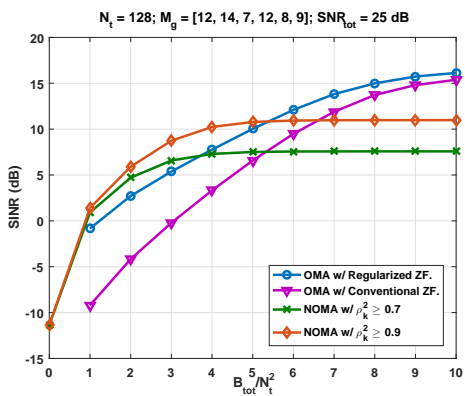}
    \caption{$K = 112$}
    \label{f10-G62}
  \end{subfigure}
  \caption{Performance for NOMA scheme from
  Algorithm~\ref{user_grouping_method} and OMA from
  Algorithm~\ref{sub_oma}.}
\label{f10-G6}
\end{figure}

Fig.~\ref{f10-G6} compares the performance of the proposed NOMA scheme
from Algorithm~\ref{user_grouping_method} with that of the OMA from
Algorithm~\ref{sub_oma} with user grouping based on $c^2_k$.  To
generate the degradation factor $c_k$ for this figure, we assume a
cell with radius of 100 meters and path-loss exponent of 2.  The
strong user for pair $k$ is uniformly placed within 50 meters from the
BS while the weak user is uniformly placed at least 60 meters away
from the BS.  Both NOMA and OMA schemes are operated with the same
system parameters. For Fig.~\ref{f10-G61}, $K = 142$, which is larger
than $N_t = 128$ while for Fig.~\ref{f10-G62}, $K = 112$.  For the
NOMA scheme, we assume a correlation coefficient between the channels
of the strong and weak users is uniform.  We consider user pairing
with 2 different thresholds on the channel correlation.  For the first
setting, users are paired only if $\rho^2_k \ge 0.7$.  For the second
setting, $\rho^2_k \ge 0.9$.  From both Figs.~\ref{f10-G61}
and~\ref{f10-G62}, we see a possible SINR gain of 3 dB when the
threshold for $\rho^2_k$ is higher.  With higher correlation between
the channels of strong and weak users in a pair, there is less
interference from weaker users due to more accurate transmit
beamforming.  Thus, user pairing is crucial for NOMA. Compared with
NOMA, OMA performs much worse when the total CDI rate is not
large. This can be attributed to more CDI vectors that need to be
quantized for OMA. Thus, with limited CDI rate, the accuracy of the
quantized CDI for OMA is worse than that for NOMA, especially when the
user load is high (Fig.~\ref{f10-G61}).  However, as $\hBtot$
increases, the SINR gap between OMA and NOMA is closing.

With fewer number of users in Fig.~\ref{f10-G62}, OMA can outperform
NOMA when the total CDI rate is large or when channel information is
highly accurate. In those regimes, the transmit beams can more
precisely cancel interference.  Additionally, there is no intra-beam
interference for OMA users to contend with.  In Fig.~\ref{f10-G62}
with $K < N_t$, we are able to add the performance of a system with
conventional zeroforcing beamforming (regularizing constant $\varphi =
0$). The optimized power and bit allocations for that system are also
obtained from Algorithm~\ref{sub_oma}.  We note that with the
optimized regularizing constant, the regularized zeroforcing performs
much better than the conventional zeroforcing does, especially when
the CDI rate is small or moderate.

From the numerical results shown in Figs.~\ref{f10-G61}
and~\ref{f10-G62}, we remark that the optimized NOMA scheme
outperforms the optimized OMA scheme when the number of CDI bits per
the number of transmit antennas squared ($\hBtot$) is between 1 and 4.
The SINR gain of NOMA over OMA increases when users in the NOMA scheme
are paired with a higher threshold on the channel correlation between
the strong and weak users.

\section{Conclusions}
\label{conclude}

We have developed a joint transmit-power, CDI-rate, user-grouping
problem for MIMO-NOMA and MIMO-OMA downlinks that maximizes the
minimum large-system SINR over all users.  For some joint power and
CDI-rate optimization, the optimal solutions can be obtained by a GP
solver.  A suboptimal progressive-filling scheme was shown to perform
close to the optimum with the computation time cut by almost 3 orders
of magnitude.  The allocation complexity can be further decreased by
decreasing the number of user groups.  Although the system SINR
increases with the number of user groups, the SINR gain is
diminishing.  For larger system SINR, users should be paired if the
correlation coefficient is sufficiently high.  Additionally, a larger
fraction of the transmit power should be allocated to user pairs.  If
the total CDI rate is low to moderate ($\hBtot$ is approximately
between 1 and 4), NOMA generally achieves a larger SINR than OMA
does. The performance gain over OMA increases when users are paired
with a high correlation threshold. On the other hand, OMA can
outperform NOMA when the total CDI-rate is very high and the
correlation threshold for user pairing is low.

In this work, the channel inaccuracy at the BS is attributed to
limited CDI from mobile users in a frequency-division duplex.  Our
results can also be applied to a time-division duplex (TDD) in which
the channel inaccuracy is caused by limited training from mobile
users.  For TDD, the number of pilots for each user pair and singleton
must be optimized instead of CDI bits.  However, if the channel
reciprocity between uplink and downlink is imperfect, the system
performance will degrade. Taking into account imperfect channel
reciprocity could be explored in the future work.  Other possible
topics include a distributed joint power and CDI-rate allocation
scheme for uplink and learning-based allocation schemes for uplink and
downlink.

\appendix

\setcounter{MYtempeqncnt}{\value{equation}}

\begin{figure*}[!b]

\normalsize

\setcounter{equation}{52}

\vspace*{4pt}

\hrulefill
\begin{equation}
  \label{gamma_LargeK2alpha0}
  \left. \gamma_{k;w}^{\infty}\right\rvert_{\alpha_k=0} =
  \frac{\zeta_g (1-\brM) (1 - 2^{-\brB_g})}{ 2\brK (c_{k}^2 - c_k^2
    \rho_{k}^2(1 - 2^{-\brB_g})) + \frac{\sigma_n^2}{P_{\text{tot}}})
    +\zeta_g (1-\brM)(1-2^{-\brB_g})}
\end{equation}

\end{figure*}

\setcounter{equation}{\value{MYtempeqncnt}}

\subsection{Proof of Proposition~\ref{prop_optpwr}}
\label{proof_prop}

From~\eqref{gamma_LargeK1} and~\eqref{gamma_LargeK2}, we see that as
$\alpha_{k}$ increases from 0, $\gamma_{k;s}^\infty$ decreases, but
$\gamma_{k;w}^\infty$ increases.  Larger $\alpha_{k}$ will lead to
allocating more transmit power to the weaker user, and less power to
the stronger user.  Hence, the optimal decay factor that maximizes the
minimum SINR of the 2 users, is at the intersection between
$\gamma_{k;s}^\infty$ and $\gamma_{k;w}^\infty$ obtained by setting
$\gamma_{k;s}^\infty = \gamma_{k;w}^\infty$.  By
substituting~\eqref{gamma_LargeK1} and~\eqref{gamma_LargeK2} into
$\gamma_{k;s}^\infty = \gamma_{k;w}^\infty$ and rearranging the
equation, we obtain the following quadratic equation
\begin{equation}
  \mathbb{A}_k (c_k^{2\alpha_k})^2 + \mathbb{B}_k c_k^{2\alpha_k} +
  \mathbb{C}_k = 0
  \label{quadeq}
\end{equation}
where coefficients $\mathbb{A}_k$, $\mathbb{B}_k$, and $\mathbb{C}_k$
are given by~\eqref{decay_a}--\eqref{decay_c}, respectively.  Since
$c_k^{2\alpha_k}$ must be greater than $0$, the only solution
for~\eqref{quadeq} is
\begin{equation}
  c_k^{2\alpha_k^*} = \frac{-\mathbb{B}_k + \sqrt{\mathbb{B}_k^2 -
      4\mathbb{A}_k \mathbb{C}_k}}{2\mathbb{A}_k} .
  \label{eqck}
\end{equation}
Note that $\mathbb{A}_k \mathbb{C}_k <0$.  Solving for $\alpha_k^*$
in~\eqref{eqck} gives~\eqref{opt_decay}.  To obtain the minimum SINR
in~\eqref{g_kmin}, we substitute $c_k^{2\alpha_k^*}$ shown above
into~\eqref{gamma_LargeK1}.

There is no intersection between $\gamma_{k;s}^\infty$ and
$\gamma_{k;w}^\infty$ for all $\alpha_k > 0$ or the intersection
occurs at $\alpha_k = 0$ if for $\alpha_k = 0$,
\begin{equation}
  \label{ineq}
  \left. \gamma_{k;w}^\infty\right\rvert_{\alpha_k = 0} \ge
  \left. \gamma_{k;s}^\infty\right\rvert_{\alpha_k = 0} .
\end{equation}
Here, $\gamma_{k;w}^\infty > \gamma_{k;s}^\infty$, $\forall
\alpha_k > 0$. Hence, the optimal $\alpha_k^* = 0$. Evaluating
$\gamma_{k;s}^\infty$ and $\gamma_{k;w}^\infty$
with~\eqref{gamma_LargeK1} and~\eqref{gamma_LargeK2}, respectively at
$\alpha_k = 0$, we obtain
\begin{equation}
  \label{gamma_LargeK1alpha0}
  \left. \gamma_{k;s}^{\infty} \right\rvert_{\alpha_k=0} = \frac{\zeta_g
    (1-\brM)(1 - 2^{-\brB_g})}{2 \brK \left( 2^{-\brB_g} +
    \frac{\sigma_n^2}{P_{\text{tot}}} \right) }
\end{equation}
and~\eqref{gamma_LargeK2alpha0}.

Substitute~\eqref{gamma_LargeK1alpha0} and~\eqref{gamma_LargeK2alpha0}
into~\eqref{ineq} to obtain condition~\eqref{ieq:alp0}. The minimum
SINR is that of the stronger user in~\eqref{gamma_LargeK1alpha0}.

\bibliographystyle{IEEEtran}
\bibliography{IEEEabrv,zfnoma}

\begin{thebibliography}{10}
\providecommand{\url}[1]{#1}
\csname url@samestyle\endcsname
\providecommand{\newblock}{\relax}
\providecommand{\bibinfo}[2]{#2}
\providecommand{\BIBentrySTDinterwordspacing}{\spaceskip=0pt\relax}
\providecommand{\BIBentryALTinterwordstretchfactor}{4}
\providecommand{\BIBentryALTinterwordspacing}{\spaceskip=\fontdimen2\font plus
\BIBentryALTinterwordstretchfactor\fontdimen3\font minus
  \fontdimen4\font\relax}
\providecommand{\BIBforeignlanguage}[2]{{%
\expandafter\ifx\csname l@#1\endcsname\relax
\typeout{** WARNING: IEEEtran.bst: No hyphenation pattern has been}%
\typeout{** loaded for the language `#1'. Using the pattern for}%
\typeout{** the default language instead.}%
\else
\language=\csname l@#1\endcsname
\fi
#2}}
\providecommand{\BIBdecl}{\relax}
\BIBdecl

\bibitem{dai2018}
L.~{Dai}, B.~{Wang}, Z.~{Ding}, Z.~{Wang}, S.~{Chen}, and L.~{Hanzo}, ``A
  survey of non-orthogonal multiple access for {5G},'' \emph{{IEEE} Commun.
  Surveys Tuts.}, vol.~20, no.~3, pp. 2294--2323, 2018.

\bibitem{mzeng17}
M.~Zeng, A.~Yadav, O.~A. Dobre, G.~I. Tsiropoulos, and H.~V. Poor, ``On the sum
  rate of {MIMO-NOMA} and {MIMO-OMA} systems,'' \emph{{IEEE} Wireless Commun.
  Lett.}, vol.~6, no.~4, pp. 534--537, Aug. 2017.

\bibitem{zhang17}
D.~Zhang, Y.~Liu, Z.~Ding, Z.~Zhou, A.~Nallanathan, and T.~Sato, ``Performance
  analysis of non-regenerative massive-{MIMO-NOMA} relay systems for 5{G},''
  \emph{{IEEE} Trans. Commun.}, vol.~65, no.~11, pp. 4777--4790, Nov. 2017.

\bibitem{liu16}
Y.~Liu, G.~Pan, H.~Zhang, and M.~Song, ``On the capacity comparison between
  {MIMO-NOMA} and {MIMO-OMA},'' \emph{IEEE Access}, vol.~4, pp. 2123--2129,
  2016.

\bibitem{nain17}
G.~Nain, S.~S. Das, and A.~Chatterjee, ``Low complexity user selection with
  optimal power allocation in downlink {NOMA},'' \emph{{IEEE} Wireless Commun.
  Lett.}, vol.~7, no.~2, pp. 158--161, Apr. 2018.

\bibitem{im19}
G.~Im and J.~H. Lee, ``Outage probability for cooperative {NOMA} systems with
  imperfect {SIC} in cognitive radio networks,'' \emph{{IEEE} Commun. Lett.},
  vol.~23, no.~4, pp. 692--695, Apr. 2019.

\bibitem{bisen21}
S.~Bisen, P.~Shaik, and V.~Bhatia, ``On performance of energy harvested
  cooperative {NOMA} under imperfect {CSI} and imperfect {SIC},'' \emph{{IEEE}
  Trans. Veh. Technol.}, vol.~70, no.~9, pp. 8993--9005, Sep. 2021.

\bibitem{yang17}
Q.~Yang, H.~M. Wang, D.~W.~K. Ng, and M.~H. Lee, ``{NOMA} in downlink {SDMA}
  with limited feedback: Performance analysis and optimization,'' \emph{{IEEE}
  J. Sel. Areas Commun.}, vol.~35, no.~10, pp. 2281--2294, Oct. 2017.

\bibitem{he21}
H.~He, Y.~Liang, and S.~Li, ``Clustering algorithm based on azimuth in mmwave
  massive {MIMO-NOMA} system,'' in \emph{IEEE/CIC Int. Conf. on Commun. in
  China (ICCC Workshops)}, Xiamen, China, Jul. 2021, pp. 118--122.

\bibitem{poor17}
Z.~Ding, Y.~Liu, J.~Choi, Q.~Sun, M.~Elkashlan, C.-L. I, and H.~V. Poor,
  ``Application of non-orthogonal multiple access in {LTE} and 5{G} networks,''
  \emph{{IEEE} Commun. Mag.}, vol.~55, no.~2, pp. 185--191, Feb. 2017.

\bibitem{dhakal19}
S.~Dhakal, P.~A. Martin, and P.~J. Smith, ``{NOMA} with guaranteed weak user
  {QoS}: Design and analysis,'' \emph{IEEE Access}, vol.~7, pp.
  32\,884--32\,896, 2019.

\bibitem{ding19}
J.~Ding, J.~Cai, and C.~Yi, ``An improved coalition game approach for
  {MIMO-NOMA} clustering integrating beamforming and power allocation,''
  \emph{{IEEE} Trans. Veh. Technol.}, vol.~68, no.~2, pp. 1672--1687, Feb.
  2019.

\bibitem{xiao18}
Z.~Xiao, L.~Zhu, J.~Choi, P.~Xia, and X.~Xia, ``Joint power allocation and
  beamforming for non-orthogonal multiple access ({NOMA}) in {5G} millimeter
  wave communications,'' \emph{{IEEE} Trans. Wireless Commun.}, vol.~17, no.~5,
  pp. 2961--2974, May 2018.

\bibitem{ding17}
Z.~Ding, X.~Lei, G.~K. Karagiannidis, R.~Schober, J.~Yuan, and V.~K. Bhargava,
  ``A survey on non-orthogonal multiple access for 5{G} networks: Research
  challenges and future trends,'' \emph{{IEEE} J. Sel. Areas Commun.}, vol.~35,
  no.~10, pp. 2181--2195, Oct. 2017.

\bibitem{ali17}
S.~Ali, E.~Hossain, and D.~I. Kim, ``Non-orthogonal multiple access ({NOMA})
  for downlink multiuser {MIMO} systems: User clustering, beamforming, and
  power allocation,'' \emph{IEEE Access}, vol.~5, pp. 565--577, 2017.

\bibitem{wang21}
Q.~Wang and Z.~Wu, ``Beamforming optimization and power allocation for
  user-centric {MIMO-NOMA} {IoT} networks,'' \emph{IEEE Access}, vol.~9, pp.
  339--348, 2021.

\bibitem{jding20}
J.~Ding and J.~Cai, ``Two-side coalitional matching approach for joint
  {MIMO-NOMA} clustering and {BS} selection in multi-cell {MIMO-NOMA}
  systems,'' \emph{{IEEE} Trans. Wireless Commun.}, vol.~19, no.~3, pp.
  2006--2021, Mar. 2020.

\bibitem{kim20}
H.-R. Kim, J.~Chen, and J.~Yoon, ``Joint user clustering and beamforming in
  non-orthogonal multiple access networks,'' \emph{IEEE Access}, vol.~8, pp.
  111\,355--111\,367, 2020.

\bibitem{chen17}
X.~{Chen}, Z.~{Zhang}, C.~{Zhong}, and D.~W.~K. {Ng}, ``Exploiting
  multiple-antenna techniques for non-orthogonal multiple access,''
  \emph{{IEEE} J. Sel. Areas Commun.}, vol.~35, no.~10, pp. 2207--2220, Oct.
  2017.

\bibitem{Tang20}
Z.~Tang, L.~Sun, L.~Cao, S.~Qi, and Y.~Feng, ``Reconsidering design of
  multi-antenna {NOMA} systems with limited feedback,'' \emph{{IEEE} Trans.
  Wireless Commun.}, vol.~19, no.~3, pp. 1519--1534, Mar. 2020.

\bibitem{zhang20}
J.~{Zhang}, Y.~{Zhu}, S.~{Ma}, X.~{Li}, and K.~K. {Wong}, ``Large system
  analysis of downlink {MISO-NOMA} system via regularized zero-forcing
  precoding with imperfect {CSIT},'' \emph{{IEEE} Commun. Lett.}, vol.~24,
  no.~11, pp. 2454--2458, Nov. 2020.

\bibitem{cui18}
J.~{Cui}, Z.~{Ding}, and P.~{Fan}, ``Outage probability constrained {MIMO-NOMA}
  designs under imperfect {CSI},'' \emph{{IEEE} Trans. Wireless Commun.},
  vol.~17, no.~12, pp. 8239--8255, Dec. 2018.

\bibitem{hoang21}
T.~M. Hoang, B.~C. Nguyen, X.~N. Tran, and L.~T. Dung, ``Outage probability and
  ergodic capacity of user clustering and beamforming {MIMO-NOMA} relay system
  with imperfect {CSI} over {Nakagami}-$m$ fading channels,'' \emph{IEEE
  Systems Journal}, vol.~15, no.~2, pp. 2398--2409, Jun. 2021.

\bibitem{ojcoms20}
K.~{Mamat} and W.~{Santipach}, ``On optimizing feedback-rate allocation for
  downlink {MIMO-NOMA} with quantized {CSIT},'' \emph{IEEE Open J. Commun.
  Soc.}, vol.~1, pp. 1551--1570, 2020.

\bibitem{nandan21}
N.~Nandan, S.~Majhi, and H.-C. Wu, ``Beamforming and power optimization for
  physical layer security of {MIMO-NOMA} based {CRN} over imperfect {CSI},''
  \emph{{IEEE} Trans. Veh. Technol.}, vol.~70, no.~6, pp. 5990--6001, Jun.
  2021.

\bibitem{wan21}
X.~Wan, E.~Li, Z.~Wang, and Z.~Fan, ``Energy-efficient resource allocation for
  multicarrier {NOMA} systems with imperfect {CSI},'' in \emph{IEEE Int. Conf.
  on Electron. Info. and Commun. Technol. (ICEICT)}, Xi'an, China, Aug. 2021,
  pp. 823--827.

\bibitem{liuxia18}
X.~Liu, J.~Zhang, and S.~Cai, ``An optimal power allocation scheme in downlink
  multi-user {NOMA} beamforming system with imperfect {CSI},'' in \emph{IEEE
  Int. Conf. on Commun. Syst. (ICCS)}, Chengdu, China, Dec. 2018, pp. 99--103.

\bibitem{kota21}
K.~K. Kota and P.~Ubaidulla, ``Sum-rate maximization in {NOMA}-based mmwave
  analog beamforming under imperfect {CSI},'' in \emph{IEEE Veh. Technol. Conf.
  (VTC2021-Spring)}, Helsinki, Finland, Apr. 2021, pp. 1--7.

\bibitem{gong19}
M.-y. Gong and Z.~Yang, ``The application of antenna diversity to {NOMA} with
  statistical channel state information,'' \emph{{IEEE} Trans. Veh. Technol.},
  vol.~68, no.~4, pp. 3755--3765, Apr. 2019.

\bibitem{gao21}
Z.~Gao, A.~Liu, C.~Han, and X.~Liang, ``Sum rate maximization of massive {MIMO
  NOMA} in {LEO} satellite communication system,'' \emph{{IEEE} Wireless
  Commun. Lett.}, vol.~10, no.~8, pp. 1667--1671, Aug. 2021.

\bibitem{zhu22}
H.~Zhu, Q.~Wu, X.-J. Wu, Q.~Fan, P.~Fan, and J.~Wang, ``Decentralized power
  allocation for {MIMO-NOMA} vehicular edge computing based on deep
  reinforcement learning,'' \emph{{IEEE} Internet Things J.}, vol.~9, no.~14,
  pp. 12\,770--12\,782, Jul. 2022.

\bibitem{zeng17}
M.~Zeng, A.~Yadav, O.~A. Dobre, G.~I. Tsiropoulos, and H.~V. Poor, ``Capacity
  comparison between {MIMO-NOMA} and {MIMO-OMA} with multiple users in a
  cluster,'' \emph{{IEEE} J. Sel. Areas Commun.}, vol.~35, no.~10, pp.
  2413--2424, Oct. 2017.

\bibitem{Saito13}
Y.~{Saito}, A.~{Benjebbour}, Y.~{Kishiyama}, and T.~{Nakamura}, ``System-level
  performance evaluation of downlink non-orthogonal multiple access {(NOMA)},''
  in \emph{Proc. Int. Symp. on Personal, Indoor, and Mobile Radio Commun.
  (PIMRC)}, London, UK, Sep. 2013, pp. 611--615.

\bibitem{mimo}
W.~Santipach and M.~L. Honig, ``Capacity of a multiple-antenna fading channel
  with a quantized precoding matrix,'' \emph{{IEEE} Trans. Inf. Theory},
  vol.~55, no.~3, pp. 1218--1234, Mar. 2009.

\bibitem{dai08}
W.~Dai, Y.~Liu, B.~C. Rider, and W.~Gao, ``How many users should be turned on
  in a multi-antenna broadcast channel?'' \emph{{IEEE} J. Sel. Areas Commun.},
  vol.~26, no.~8, pp. 1526--1535, Oct. 2008.

\bibitem{gpboyd07}
S.~{Boyd}, S.~{Kim}, L.~{Vandenberghe}, and A.~{Hassibi}, ``A tutorial on
  geometric programming,'' \emph{Optimization and Engineering}, vol.~8, no.~67,
  pp. 67--127, Apr. 2007.

\bibitem{chassein14}
\BIBentryALTinterwordspacing
A.~Chassein and M.~Goerigk, ``Robust geometric programming is co-{NP} hard,''
  Dec. 2014, unpublished. [Online]. Available:
  \url{http://nbn-resolving.de/urn:nbn:de:hbz:386-kluedo-39380}
\BIBentrySTDinterwordspacing

\bibitem{mollah18}
M.~A. Mollah, X.~Yuan, S.~Pakin, and M.~Lang, ``Rapid calculation of max-min
  fair rates for multi-commodity flows in fat-tree networks,'' \emph{{IEEE}
  Trans. Parallel Distrib. Syst.}, vol.~29, no.~1, pp. 156--168, Jan. 2018.

\bibitem{wagner12}
S.~Wagner, R.~Couillet, M.~Debbah, and D.~T.~M. Slock, ``Large system analysis
  of linear precoding in correlated {MISO} broadcast channels under limited
  feedback,'' \emph{{IEEE} Trans. Inf. Theory}, vol.~58, no.~7, pp. 4509--4537,
  Jul. 2012.

\end{thebibliography}

\begin{IEEEbiography}[{\includegraphics[width=1in,height=1.25in,clip,keepaspectratio]{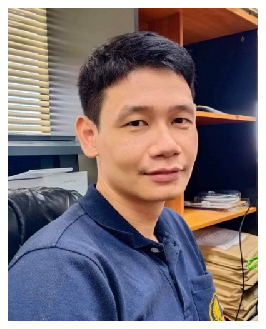}}]{Kritsada Mamat} was born in Suphanburi, Thailand in 1984.  He received the
B.Eng. degree in electrical engineering from Burapha University,
Chonburi, Thailand, in 2007, and the M.Eng. and D.Eng. degrees in
electrical engineering from Kasetsart University, Bangkok, Thailand,
in 2010 and 2016, respectively.  After graduation, he joined the
Department of Electrical Engineering, Faculty of Engineering,
Kasetsart University as a Post-Doctoral Researcher for one year.  He
is currently an Assistant Professor with the Department of Electronic
Engineering Technology, College of Industrial Technology, King
Mongkut's University of Technology North Bangkok, Bangkok,
Thailand. His research interest is in optimization techniques for
wireless communication channels.
\end{IEEEbiography}

\begin{IEEEbiography}[{\includegraphics[width=1in,height=1.25in,clip,keepaspectratio]{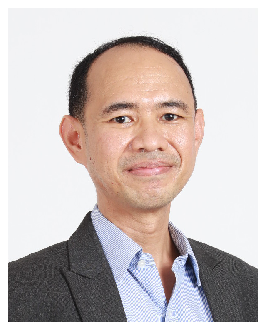}}]{Wiroonsak Santipach} (Senior Member, IEEE)
received the B.S. (summa cum laude), M.S., and Ph.D. degrees all in
electrical engineering from Northwestern University, Evanston,
Illinois, USA, in 2000, 2001, and 2006, respectively. In 2006, he
joined the Department of Electrical Engineering, Faculty of
Engineering, Kasetsart University, Bangkok, Thailand, as a Lecturer
and since 2019, he has been a Professor.  In 2013 and 2019, he was a
visiting scholar at the Institute of Telecommunications, Technische
Universit\"{a}t Wien, Austria.  He has authored 2 books titled
Introduction to Telecommunication Engineering (Kasetsart University
Press, 2016) and Wireless Communications With Limited Feedback
(Kasetsart University Press, 2020), and coauthored more than 40
technical papers.  His current interest includes limited feedback in
NOMA and MIMO channels, and deep learning in wireless channels.
\end{IEEEbiography}

\end{document}